\documentclass[12pt,preprint]{aastex}

\slugcomment{To appear in A. J., Vol. 127 (2004 January)}

\shorttitle{TV COLUMBAE}
\shortauthors{Rana et al.}

\begin{document}


\title{Periodicities in the X-Ray Intensity Variations of TV Columbae: An Intermediate Polar}


\author{V. R. Rana\altaffilmark{1} and K. P. Singh}
\affil{Department of Astronomy \& Astrophysics, Tata Institute of Fundamental Research, Homi Bhabha Road, Mumbai 400~005, INDIA}
\email{vrana@tifr.res.in, singh@tifr.res.in}

\author{E. M. Schlegel}
\affil{Harvard-Smithsonian Center for Astrophysics, 60 Garden Street, Cambridge, MA 02138, U.S.A.}
\email{ems@head-cfa.harvard.edu}

\and

\author{P. Barrett}
\affil{Space Telescope Science Institute, ESS/Science Software Group, Baltimore, MD 21218, U.S.A.}
\email{barrett@stsci.edu}


\altaffiltext{1}{Joint Astronomy Programme, Department of Physics, Indian Institute of Science, Bangalore 560~012, INDIA}

\begin{abstract}
We present results from a temporal analysis of the longest
and the most sensitive X-ray observations of TV~Columbae -- a cataclysmic
variable classified as an intermediate polar. The observations were carried
out with the $Rossi X-Ray Timing Explorer (RXTE)$ Proportional Counter Array, 
$ROSAT$ Position Sensitive Proportional Counter, and $ASCA$. 
Data were analyzed using a one-dimensional CLEAN and Bayesian algorithms.
The presence of a nearly sinusoidal modulation due to the spin of the white
dwarf is seen clearly in all the data,
confirming the previous reports based on the $EXOSAT$ data.
An improved period of $1909.7\pm2.5$s is derived for the spin from the 
$RXTE$ data.
The binary period of 5.5hr is detected unambiguously in X-rays 
for the first time.  
Several side-bands due to the interaction of these periods are observed in
the power spectra, thereby suggesting contributions from
both the disk-fed and the stream-fed accretion for TV~Col.
The accretion disk could perhaps be precessing as side-bands due to the
influence of 4 day period  on the orbital period are seen.
The presence of a significant power at certain side-bands of the spin
frequency indicates that the emission poles are asymmetrically located.
The strong power at the orbital side-bands seen in both the $RXTE$ and
$ROSAT$ data gives an indication for an absorption site fixed in the
orbital frame.  
Both the spin and the binary modulation are found to be energy- dependent. 
The fact that the spin modulation amplitude in TV~Col decreases 
with energy is confirmed.
Hardness ratio variations and the energy dependent modulation depth 
during the spin modulation can be explained by partially covered
absorbers in the path of X-ray emission region in the accretion stream.
Increased hardness ratio during a broad dip in the intensity
at binary phase of 0.75--1.0 confirms the presence of a strong attenuation 
due to additional absorbers probably
from an impact site of the accretion stream with the disk or magnetosphere.
\end{abstract}


\section{Introduction}

The cataclysmic variable (CV), TV~Columbae, is classified as an 
intermediate polar (IP) containing a magnetic white dwarf accreting 
from the Roche-Lobe of a late-type
dwarf companion via an accretion disk and an accretion channel.
It was identified by Charles et al. (1979) as the optical counterpart of a hard
X-ray source (2A 0526-328) first discovered with the $Ariel~5$ satellite
(Cooke et al. 1978).
An optical star with V$\sim$13--14 mag, TV~Col is at a distance of
$368^{+17}_{-15}$ pc (McArthur et al. 2001). 

A 1911$\pm$5 s period has been detected from this source in X-rays using the
$EXOSAT$ observatory (Schrijver, Brinkman, \& van der Woerd 1987).  The X-ray period has
been identified as the spin period of the white dwarf, and the spin-phased 
light curves have been found to be energy dependent (Norton \& Watson 1989).
X-ray observations of TV~Col have also been reported from the
$ROSAT$, $ASCA$, and $Ginga$ satellites by 
Vrtilek et al. (1996), Ezuka \& Ishida (1999), and Ishida \& Fujimoto (1995), 
respectively, but they mainly discuss the spectral
characteristics of the source rather than the timing properties.
Vrtilek et al. (1996), however, reported a failure to detect the
X-ray spin period in the $ROSAT$ observation analyzed by them.
A period of 1938$\pm$10s was claimed to be present in the optical bands
(UBV) at a
level of $\sim$6\% (full amplitude) by Bonnet-Bidaud, Motch, \& Mouchet (1985)
(see their note added in proof) after the discovery of 32 minute 
period in X-rays with the $EXOSAT$ satellite was reported by 
Schrijver et al. (1985).

TV~Col also exhibits additional periods of 5.2 hr, 5.5 hr and 4 days 
(Motch 1981; Hutchings et al. 1981;
Bonnet-Bidaud et al. 1985; Barrett, O'Donoghue, \& Warner 
1988; Hellier, Mason, \& Mittaz 1991;
Hellier 1993; Augusteijn et al. 1994) observed in optical photometry 
and spectroscopy, which add further complexities.
None of these periods have been seen convincingly in the X-ray observations
reported so far, although Hellier, Garlick, \& Mason (1993) did report
some evidence for intensity dips separated by an orbital cycle in the
data taken with the $EXOSAT$ observatory.
The spectroscopic period of 5.5 hr is believed to represent the
binary orbital modulation and shows partial eclipses.  The 4-day period
is possibly due to the nodal precession of the accretion disk around the white
dwarf (Barrett et al. 1988 and Hellier 1993).
The 4-day period was also reported from UV observations by Mateo, Szkody,
\& Hutchings (1985).
The 5.2 hr period likely represents the beat between
the 5.5 hr and 4-day periods.
Recently Retter et al. (2003) have reported another period at 6.3 hr
in the optical data, and identified it as representing the permanent positive
superhump period of the binary system.
The positive superhump is usually explained as the beat between the binary
period and the precession of an eccentric accretion disk in the apsidal plane
(see reviews by Patterson 2001; O'Donoghue 2000).

The power spectrum of IPs in X-ray band can be very complex containing the
fundamental system frequencies, their harmonics and sometimes the 
side-band frequencies
due to beating of the fundamental periods and their harmonics.
The X-rays, which are believed to originate from the shock
heated plasma region near the surface of the white dwarf, show modulation
at all or some of the above mentioned frequency components. 
Hence a long and well
sampled hard X-ray data particularly taken from an instrument with large
effective area and high
time resolution can provide important clues about the mode of accretion
and the geometry of these sources.

Here, we report on the longest and the most sensitive broad-band
X-ray observations carried out in the hard X-rays with the 
$Rossi X-Ray Timing Explorer (RXTE)$, and in the soft X-rays 
with the $ROSAT$ satellites. 
Although these observations are not simultaneous these are 
the best available for carrying out temporal analysis to search 
for various periodicities in the X-ray emission from TV~Col and to
study their energy dependence. 
Results from the timing analysis of simultaneous soft and hard 
X-ray observations with the $ASCA$ satellite are also presented.
A detailed timing analysis of these observations has not been
reported so far except for a preliminary report in Singh et al. (2003).
The paper is organized as follows: In the next section
we present details of observations and data reduction in the hard X-ray and
the soft X-ray energy bands. Section 3 contains results obtained from the power
spectral analysis, and the light curves folded on the spin period and
the orbital period of the system, followed by a discussion and interpretation
of results in \S 4.  Finally, we present a summary of our results and
the conclusions in \S 5.

\section{Observations and Data Reduction}

\subsection{Hard X-rays}

TV~Col was observed with the Proportional Counter Array (PCA) onboard the
$RXTE$ during the years 1996 (PI: E. Schlegel) and 2000 (PI: K. Mukai).
The PCA consists of five Proportional Counter Units (PCUs),
each filled with xenon gas and split into three layers, plus a propane
filled veto layer. Each PCU has a large
collecting area of 1300 $cm^2$, energy bandwidth of 2--60 keV and
a collimator with a field of view of 1$^{\circ}$ (FWHM).
The energy resolution of a PCU is $\sim$ 18\% at 6 keV.
More details about the PCA can be found in Jahoda et al. (1996).

The complete observation log is given in Table 1.
Columns (1) and (2) list the start and end times of the observation.
The name of the satellite is given in column (3) and the effective exposure
times are given in column (4).
The mean source count rates in 2--20 keV energy band for the $RXTE$, 
0.1--2.2 keV energy band for the $ROSAT$, and 0.7--10 keV energy band for the
$ASCA$,
after background subtraction (see below), are given in column (5),
followed by comments on the frequency and duration of each exposure
in column (6).
All the PCUs were ``ON'' during most of the 1996 observations whereas only 
three PCUs were ``ON'' during the 2000 observations.
Of the total exposure time of 79 ks in 1996, the time for which less 
than five PCUs were ``ON'' was $\sim$12 ks.
The count rates shown are
scaled to the same number of PCUs assuming identical area for each PCU.

The PCA data were processed using the standard programs of the software
package FTOOLS 5.1\altaffilmark{2}. 
\altaffiltext{2}{the information about FTOOLS can be found at 
{\it http://heasarc.gsfc.nasa.gov/ docs/corp/software.html}}.
Both standard 1 (time resolution of 1 s) and standard 2
(time resolution of 16 s) data were extracted and analyzed.

X-ray light curves (source plus background) were obtained using the
following selection criteria: we use only the `good time interval' (GTI)
data when extracting the light curves using the SAEXTRCT program
in FTOOLS.  The GTI selects data when the Earth elevation angle is
greater than $10^{\circ}$ and the offset between the pointing
direction of the satellite and the centre of the field-of-view is less
than $0^{\circ}$.02.  The times of high
concentration of electrons that increase the noise at low energies
were excluded.
The latest set of combined background models ``CM L7
Linear'' for faint sources were used to create the
background light curves for each observation.
The background light curves were then subtracted
from the (source plus background) light curves to produce the
X-ray light curves for the source.

We analyzed the standard 2 data from only the top xenon
layer of the PCA for a better signal-to-noise ratio than is obtainable
from the three layers combined.  Light curves
in the energy band of 2--20 keV were thus extracted and analyzed.
The mean normalized count rate was found to be nearly the same in the 1996
and the 2000 observations (see Table 1).   The normalized X-ray light
curve in the 2--20 keV energy band with a bin time of 64~s is shown in
Figure 1 for the
1996 observations. The observations done in the year 2000 suffered from
poor sampling
(see Table 1), and are therefore not analyzed any further.

\subsection{Soft X-rays}

TV~Col was observed thrice with the $ROSAT$ (Tr\"{u}mper 1983)
Position Sensitive Proportional Counter (PSPC) during 1991--1993
(PI: S. D. Vrtilek for 1991 observation and P. Barrett for 1992 and 1993
observations).  The observation log is given in Table 1.
The X-ray image and the light curves were extracted
using the XSELECT program in the FTOOLS.
The soft X-ray light curve of TV~Col was obtained by
extracting the counts with a resolution of 4~s from the unsmoothed
PSPC images using a circle with a radius of $3\arcmin$.75 centered on
the peak position.
The background was estimated from several neighboring source free regions and
found to be steady with a count rate of 0.041 counts s$^{-1}$.
The observed count rate, after background subtraction,
varies between 0.2 and 0.7 counts s$^{-1}$
during each of the three observations.
There are no confusing sources in the neighborhood, the
next brightest source in the vicinity is about 10 times fainter and
$10.5\arcmin$ away.
We have performed a detailed analysis of the X-ray data from the
longest observation with an exposure time of 22557~s obtained during
1993 February 9--11.
The source X-ray light curve in the 0.1--2.2 keV energy band with a bin time
of 128~s is shown in Figure 2 for the 1993 observation.
The two observations in 1991 and 1992 were very short and infrequent
(see Table 1), and are therefore not analyzed any further.
Analysis of the 1991 observation was presented by
Vrtilek et al. (1996).

In addition to the analysis of the above soft and hard X-ray data separated by
3 yr, we have also analyzed a simultaneously obtained 0.7--10 keV data from 
the $ASCA$ SIS0 (Yamashita et al. 1997) observations during 1995 February 28
to March 01. (See Table 1 for details). We  analyzed SIS0 `bright2' mode 
data with a time resolution of 16~s. We applied a strict screening criteria 
for the parameter ``RBM-CONT'' setting its value less than 75,
instead of the standard screening. This essentially removes the excess count
rate due to a flare mainly in the soft X-rays (below 2 keV).  
Source photons were extracted from a circular region with a radius 
of $4\arcmin$ centered on the peak position. 
The background light curves were extracted from the neighboring CCD chip 
which is free from the source counts. 
The observed count rate, after the background subtraction, varies between 0.3
and 0.9 counts s$^{-1}$ with an average value of $\sim$0.6 counts s$^{-1}$.
The spectra corresponding to the source and the 
background were also
extracted from the corresponding regions. Results from a detailed spectral 
analysis of these data were reported by Ezuka \& Ishida (1999).

In this paper we will mainly concentrate on the analysis of
the $RXTE$ and the $ROSAT$ data, using the $ASCA$ data only to verify
some of the results.

\section{Analysis and Results}

\subsection{Power Spectra}

Of the six observations listed in Table 1, only three observations, namely the
1993 ($ROSAT$), the 1996 ($RXTE$), and the 1995 ($ASCA$),
are sufficiently long and with good sampling (i.e. more exposure
and less gap between them) to provide adequate
frequency resolution to
search for the fundamental periods, their harmonics and side-band
components in the power spectral distribution of the X-ray
light curves of TV~Col.
The frequency resolution is best for the 1996 $RXTE$ data with a value
of $6.8184 \times 10^{-7}$ s$^{-1}$.  The 1993 $ROSAT$ data and the 1995 $ASCA$ 
data have frequency resolutions of $2.0605 \times 10^{-6}$ s$^{-1}$ and
$2.6179 \times 10^{-6}$ s$^{-1}$, respectively.
All observations from the $ROSAT$, the $RXTE$,
and the $ASCA$ satellites have gaps due to occultation of the source by the
Earth and the times when the instruments are switched off during
the passage of the satellites through regions of high particle
background  like South Atlantic Anomaly.  This leads to
complications in the power spectrum as the true variations in the
source are further modulated by the irregular and infrequent sampling
defined by the window function of the data.
Therefore, the ``dirty'' power density spectrum first calculated from the
light curve was
followed by the deconvolution of the ``window'' function from the data
using the one-dimensional ``CLEAN'' algorithm as implemented in the
PERIOD program (Version 5.1) available in the STARLINK
package  (see Roberts, Lehar, \& Dreher 1987 for details on the CLEAN method).
We applied the bary-centric correction to all the data sets before
carrying out the power spectral analysis.

The dirty power spectrum obtained from the 1996 $RXTE$ observation in the
2--20 keV energy band is shown in Figure 3 $(top)$. The
corresponding window power spectrum is shown plotted in the middle panel.
The CLEANed power
spectrum after deconvolution of the window power is presented
in the bottom panel of the figure.
The power plotted (here and below) is square of half the amplitude of the
corresponding sinusoidal component at that frequency.
Similarly, the CLEANed power spectrum obtained from the 1993 $ROSAT$
observation in the energy band of 0.1--2.2 keV is shown in Figure 4
along with its dirty power and the window power. The CLEANed
power spectra presented here
were obtained after 1000 iterations of the CLEAN procedure with a loop gain of
0.1. A large value of CLEAN iterations
with a small value of loop gain result in a stable power spectrum with
significant noise reduction (Norton et al. 1992a, 1992b).
Excessive cleaning can, however, result in a considerable reduction of
the amplitude of the real peak.
We chose the optimum values of the above two parameters such that the reduction
in the amplitude of real peak is minimum and the reduction in the noise is maximum.
Since we could get the direct measurement of amplitude from the spin-folded
light curves, we used it as a control for deciding the number of iterations
and gain values.
After 1000 iterations the amplitude of the peak corresponding to the
spin period of the system reduced by only $\sim$15\% (typical in such
applications, e.g., Norton, Beardmore, \& Taylor 1996).
Several peaks can be seen in the power spectra shown in Figs. 3 \& 4.
Peaks seen in a power spectrum can correspond both to actual 
frequencies of the system and to the noise present in the data.
In order to determine which of the peaks are because of the system, we had
to estimate the noise level in the power spectra.
To estimate the noise level, a procedure similar to the one described in
Norton et al. (1996) was followed. The absolute amplitude of the power
level above
which at least N/10 number of noise peaks are present, where N is the
number of independent frequency samples, was estimated. The value of N
can vary according
to the length of observation and the time resolution of the light curve
from which the power spectrum has been obtained. This estimated value of noise
level corresponds to 90\% confidence limit, since there is only 10\%
chance that the peak above this level is due to noise. The noise level
determined this way is given in Table 2, and the peaks
above this level are considered to be real. The noise in the CLEANed 
power spectra
estimated this way is nearly constant in the frequency range of interest.

In order to further confirm whether the peaks present in the
CLEANed power spectra above the noise level are indeed related to the system,
an independent method namely the ``Bayesian'' method was adopted.
This method is particularly useful for the detection of a periodic signal in
the data when there is no prior knowledge about the presence of such a signal
or its characteristics. It is also equally sensitive to sinusoidal and
non-sinusoidal modulations, since it does not require a prior information about
the shape of the signal. This
approach is mathematically more rigorous and can provide an optimum number of
frequencies allowed by data with a better accuracy. Detailed information
about this method can be found in Bretthorst (1988) and the comparison of
this method with the other more commonly used methods like Fast Fourier
Transform 
and epoch folding is presented in Gregory \& Loredo (1996). The Bayesian
method is an iterative method in which a periodogram of the log of
Student t-distribution versus frequencies is generated. Student t-distribution
is essentially
a posterior probability density that a frequency $\omega$ is present
in the data when no prior information about it is available. The dominant
period corresponding to the highest probability is then identified in
the periodogram and made as a part of the model by specifying its period.
This essentially removes that period from the light curve. This
procedure is repeated for the other dominant periods in the periodogram.
The periodogram obtained using this algorithm also showed two dominant
peaks corresponding to the spin period and the orbital period of the system,
apart from other frequency components.
Several peaks present in the CLEANed power spectrum were thus identified using
the Bayesian algorithm.
This list of identified components for TV~Col was then compared with the
list of expected components (Table 1 of Norton et al. 1996).
The absolute power corresponding to various identified system components 
are listed in Table 2.
In order to match the observed frequency components with the expected
components, we calculated a 90\% confidence interval, for each of the observed
peak centroid by fitting a Gaussian model. If the expected component is within
this 90\% confidence interval, we considered it to be a match.
Apart from these identified components there are
several other unidentified peaks with power well above
the noise level but do not match with any of the system frequency. 
In Table 2, we have listed the frequency components that are identified
with the system components and are confirmed using the Bayesian method. 
Strong but unidentified peaks are also listed in Table 2. Of these about 
65\% were confirmed using the Bayesian method.
Only the identified frequency
components are marked in Figures 3 and 4. In the rest
of the paper, however, we will discuss the power spectra obtained using the
CLEAN method but focus only on the frequencies confirmed using the
Bayesian method.

The power spectra shown in
Figs. 3 \& 4 show a dominant peak at a frequency, $\omega$, corresponding
to a period of 1909.7$\pm$2.5~s 
derived from the 1996 $RXTE$ data, and 1903.3$\pm$7.5~s from the 1993 $ROSAT$
data. The strength of the peak at these periods suggests that $\omega$ is 
very likely to be the spin frequency of the white dwarf in TV~Col.
The errors on these periods are derived by calculating the half-bin size of 
a single frequency bin, centered on the peak present in the periodogram. 
These values are consistent
with the previously reported value of 1911$\pm$5~s (Schrijver et al. 1987;
Norton \& Watson 1989) from the $EXOSAT$ data.
Significant power is also observed at periods
corresponding to a value of 19819$\pm$267~s from the 1996 $RXTE$,
and 19413$\pm$770~s from the 1993 $ROSAT$ data, consistent with the 5.5~hr
(19800~s) orbital period (The corresponding orbital frequency is represented
here by $\Omega_0$). This period is detected without any ambiguity in $RXTE$
data, but is merged with other nearby components in the $ROSAT$ data.
Since both the data sets show the presence of $\Omega_0$ component, a
strong orbital modulation is expected to be present in the data.
Several side-band frequencies due to interactions among these periods
are also seen.
Power at the side-band frequencies $\omega+\Omega_0$, $\omega+2\Omega_0$,
$\omega+3\Omega_0$, 2$\omega+3\Omega_0$ and 3$\omega-2\Omega_0$
are seen above the noise.
A modulation at the side-bands due to the interaction of the orbital
and the 4 day period due to the precession of disk (precession
frequency,--$\Omega_{pr}$) is also observed.
The value adopted for the precession frequency is $2.8935 \times 10^{-6}$
s$^{-1}$ (Barrett et al. 1988; Hellier 1993).
Power seen at a frequency component corresponding to
the 5.2h period is identified with the side-band frequency corresponding
to  $\Omega_0+\Omega_{pr}$ in the $RXTE$ data (in 2--5 keV energy band),
but is not resolvable in the $ROSAT$ data.
Various other side-bands corresponding to the first, second and third 
harmonics of $\Omega_0$ and $\Omega_{pr}$ are also present in Figs. 3 \& 4. 
No significant power is seen, however, at the previously reported
superhump period of 6.3~hr seen in optical photometry by Retter et al. (2003).
The X-ray power level for this period has been shown with a vertical
dotted line in Fig. 3.

We also extracted a CLEANed power spectrum from the 1995 $ASCA$ SIS0 
observations 
of TV~Col in the 0.7--10 keV energy band, using the same values for the 
number of CLEAN iterations and the loop gain as used for
generating the $RXTE$ and the $ROSAT$ power spectra. The power
spectrum shows a strong peak at a period of 1909.9$\pm$9.5~s,consistent with
the value of $\omega$ derived from the $RXTE$ and the $ROSAT$ data.
A prominent peak is also observed at a period of 19099$\pm$957~s, 
consistent with the 5.5~hr binary period.

\subsection{Energy Dependence of Power Spectrum}

We have also studied the effect of different energy bands on the power spectrum
and obtained power spectra for three different energy bands, viz., 2--5 keV,
5--10 keV, and 10--20 keV using the $RXTE$ data, and two softer energy
bands of 0.1--0.75 keV and 0.75--2.2 keV from the $ROSAT$ data.
These energy resolved CLEANed power spectra are displayed in Figures 5 \& 6,
for the $RXTE$ and the $ROSAT$ bands respectively.
The absolute power of the various identified frequency components in
these energy resolved power spectra are listed in
Table 2. The peak power
corresponding to  $\Omega_0$ decreases with energy and is near the
noise level in the highest energy band (see Fig. 5).
However, some of the side-band frequencies for the orbital period seem to be
present at higher energies, indicating that the effect of orbital
modulation may still
be present. On the other hand the peak power
for $\Omega_0$ in the softest energy band of 0.1--0.75 keV becomes comparable
with that for $\omega$.
According to the theoretical model of Norton et al. (1996) the amplitude at
the frequency component $\omega+3\Omega_0$ is predicted to be zero.
We observed a power at
this component which is above the noise level in the high energy
bands (greater than 2 keV) but is not detected in the soft energy bands 
(less than 2 keV).
We also observed a little power above the noise level corresponding to 
the second harmonic (3$\Omega_0$) of the orbital period above 2 keV energy
bands.

As listed in Table 2, it is mostly the positive side-band
frequencies due to the interaction of $\omega$ and $\Omega_0$ that are present
in the power spectra of TV~Col, the corresponding negative side-band
frequencies being absent.
The only component present corresponding to the
negative side-band frequency is $3\omega$-$2\Omega_0$ in $ROSAT$ soft X-ray
energy bands. 
The energy resolved power spectra also shows the presence of several 
side-band frequency components due to $\Omega_0$ and $\Omega_{pr}$ and
their harmonics (Table 2).

\subsection{Spin Modulated Light Curves}

We have used the spin period derived by us using the 1996 $RXTE$ data,
since this is the most accurate determination so far, to fold all the 
data sets analyzed in this paper.
The folded light curves in the three energy bands
of $RXTE$ are shown in the top three panels of
Figure 7. The spin-folded light curve in the $ROSAT$ soft X-ray energy band
of 0.1--2.2 keV is shown in Figure 8 $(top)$.
Similarly, the spin-folded light curves for the $ASCA$ energy bands of
0.7--2 keV and 2--10 keV are shown in the top two panels of Figure 9.
Our attempts to obtain a new ephemeris for the spin period of
TV~Col were not successful since TV~Col
is a highly variable source and the shape profile of its spin-modulated
light curve (see Figs. 1 \& 2)
changes from cycle to cycle making it very difficult to identify
a permanent feature with respect to which we can define the ephemeris.
Therefore, an arbitrarily defined epoch, HJD=2,450,304.5 was used here to fold
all data with the spin period.
The modulation is nearly sinusoidal in all the energy bands 
(see Figs. 7, 8 \& 9) with the
softer energy bands showing a relatively higher amplitude than the
harder bands.  The relative modulation amplitudes, defined as the ratio
of the amplitude of the sine wave to the average intensity, are about 
20\%$\pm$2\%, 14\%$\pm$0.3\%, 8.5\%$\pm$0.2\% and 6\%$\pm$0.5\% of the mean
intensity in the energy bands of 0.1--2.2 keV, 2--5 keV, 5--10 keV, and 
10--20 keV respectively.  
For the two $ASCA$ energy bands of 0.7--2 keV and 2--10 keV, 
the modulation amplitudes are $\sim$17.5\%$\pm$2.5\% and 9\%$\pm$2\% 
respectively.
The modulation in the $ROSAT$ band appears nearly out
of phase by 180$^\circ$ compared with the hard $RXTE$ bands which vary in
phase with each other.   
However, the spin modulations in the $ASCA$ energy bands of 0.7--2 keV
and 2--10 keV are observed to be in phase with each other.
The shape of the spin-folded light curve is also found to be energy 
dependent (see Figs. 7, 8 \& 9) thus confirming the previous results upto
10 keV by Norton \& Watson (1989).  The shape changes
from a sinusoid to more like a flat-topped profile with increasing
energy. 

Hardness ratio curves obtained from the three energy bands of $RXTE$ are shown
plotted in the bottom two panels of the Fig. 7 as a function of the
spin phase.   Two hardness ratios, HR1 and HR2, are defined for the
$RXTE$ data as follows:- HR1 is the ratio of the
count rate in 5--10 keV to count rate in 2--5 keV energy bands, and HR2 is
the ratio of the count rate in 10--20 keV to count rate in 5--10 keV energy
bands. A third hardness ratio, HR3 is defined as the ratio of the count rate in
0.75--2.2 keV energy band to the count rate in 0.1--0.75 keV energy band
for the $ROSAT$ data and is shown plotted in  
Fig. 8$(bottom)$ as a function of the spin phase.
The hardness ratio HR4, for the $ASCA$ energy bands is defined
as the ratio of the count rate in 2--10 keV energy band to the count rate in 
0.7--2 keV energy band and is shown plotted in Fig. 9$(bottom)$.
HR1 shows a strong modulation being 180$^\circ$ out of phase with
the intensity
modulation i.e., the maximum in HR1 occurs at the lowest intensity (spin phase = 0.5).
A similar variation can also be seen in the HR2 but is less modulated.
Hardness ratio, HR3 hardly shows any effect due to spin modulation and remains
constant over the entire spin cycle.
The hardness ratio curve HR4 also does not show any significant 
variation over the
entire spin cycle, except for a small change during the intensity minimum
but with large error bars due to the low count rate in this phase range.

\subsection{Binary Period Modulated Light Curves}

We have used the most accurate ephemeris available for the orbital period
as given by Augusteijn et al. (1994):

T(HJD) = 2,447,151.2324(11) + 0.22859884(77) N,      \\
to fold data from all the three satellites. 
The folded light curves
in four energy bands from soft ($ROSAT$) to hard ($RXTE$) energies
are shown in Figure 10.
Similar folded light curves for the $ASCA$ energy bands are plotted
in the top two panels of Figure 11.
The effect of orbital modulation is seen as a large asymmetric dip
which is best seen in the 2--5 keV energy band.
This dip is broad, spanning the phase range of 0.75--1.2 in the
soft X-ray energy band.
The main dip splits into two because of the presence of a small interpulse
centered at phase 0.86.
The two minima of these small dips occur at orbital phases of $\sim$
0.80 and $\sim$0.93, respectively.
The width of these low-intensity features appear to decrease with
increasing energy -- being the broadest in the $ROSAT$ 0.1--2.2 keV
soft X-ray band,
and visible at the highest energies only during the
orbital phases of 0.75--1.0.
Unfortunately, the data could not cover the complete orbital phase cycle
during the 1993 $ROSAT$ observations and there is a data gap in the 0.8--0.9
phase range (see Fig. 10, $top$).
This is, however, compensated by the $ASCA$ observation in the 0.7--2 keV
energy band that shows a clear dip during the 0.8--0.9 phase range.
Persistent X-ray emission is seen during these low intensity
phases.

Hardness ratios defined in \S 3.3 have been plotted as a function of the orbital
phase and shown in the bottom panels of Figure 11 (HR4) and in 
Figure 12 (HR1 to HR3).
HR1 shows a strong increase during the minimum intensity
(orbital phase = 0.7--1.0).
The shape profile of this broad peak in HR1 is exactly opposite but similar
in size to the broad dip seen in the folded light curve of Fig. 10.
A similar effect is also visible in the HR2.
Hardness ratio, HR3 also shows a small increase at phase 0.95, present
in HR1 and HR2, however the error bars are comparatively larger here
because of the low count rate during intensity minimum in the $ROSAT$ 
bands in this phase range.
Hardness ratio HR4 also shows a strong increase during the minimum intensity phase
and remains almost constant during the rest of the binary cycle. The error bars during
the phase range 0.8--0.9 are very large here because of the small 
count rate in this phase range.

\section{Discussion}

The longest ever observed X-ray light curves of TV~Col in various 
energy bands ranging from the very soft (0.1 keV) to hard (20 keV) X-rays 
have been analyzed in detail.
The light curves have the best timing resolution and the highest
sensitivity at the highest energies reported so far.
A variety of behavior and complexity is observed in these light curves. 
Power density spectra obtained using the CLEAN and the Bayesian 
methods show strong peaks at two fundamental periods i.e., 
the spin period and the orbital
period of the system in both the soft X-ray
and the hard X-ray bands indicating a strong modulation at these periods.
In general, the X-ray modulation at the spin period can arise due 
to the occultation of the X-ray emitting source by the white dwarf body 
and/or due to photoelectric absorption and electron scattering 
in the accretion curtain.
On the other hand, the most common cause for the orbital modulation in
X-ray light curves is due to attenuation of X-rays from material
fixed in the orbital frame.
The results presented in \S 3 are examined and discussed below within 
the above framework.

\subsection{Power Spectra}

Various frequency components corresponding to the interaction of three
fundamental periods (spin, orbital and precession) and side-bands due to
their harmonics are present in the power density spectra of TV~Col.
Norton et al. (1996) have predicted various frequency components that
can be present in a power spectrum of an IP assuming some generally
known sites for emission and absorption components.
The presence of several components in the power spectra of TV~Col suggest
contribution from several emission and attenuation sites that are listed
by Norton et al. (1996).
Frequency components in TV~Col are better resolved in the 1996
$RXTE$ data because
of its better sampling and long length of observation compared with the 1993
$ROSAT$ observation. Several of these components are merged with each other
in the 1993 $ROSAT$ data, especially at low frequencies, and so is the case
with the 1995 $ASCA$ data as well.
Significant power at frequency components corresponding to
$\omega$, $\Omega_0$, $\omega+\Omega_0$, $\omega$+2$\Omega_0$ and
3$\omega$-2$\Omega_0$ in several energy bands,
predict that TV~Col is an asymmetric system in which both the emission poles
are not diametrically opposite and have slightly different emission properties
(see Norton et al. 1996).
The amount of asymmetry depends on whether the signal corresponding to
the rotation of the asymmetrically placed poles is strong or weak.
The presence of an additional source of emission due to the impact of
the accretion stream with the accretion disk or magnetosphere is also
predicted. A strong signal at the orbital period below 10 keV energy
clearly indicates an explicit
orbital modulation that is most likely due to the attenuation
introduced by the impact site (see below \S 4.3).
In addition, the influence of the 4 day precession period has been observed in term of
its various side-band frequencies with the orbital period.

The absence of power corresponding to the frequency at the side-band
$\omega-\Omega_0$ and presence of dominating power at $\omega$ suggest
a significant contribution from the disk-fed accretion for TV~Col.
According to Wynn \& King (1992), the power spectrum of an IP with a simple 
diskless geometry is dominated by signals at frequency 
components $\omega-\Omega_0$ or 2$\omega-\Omega_0$. 
The absence of dominating power
at these frequency components in TV~Col suggests a predominant  
contribution from the disk-fed accretion component.
It is, however, very unlikely that the accretion is taking place
entirely via the disk because of the presence of various other side-band 
frequency components.
Signal at $\omega+\Omega_0$ is seen in two energy bands
(see Table 2) indicating amplitude modulation of the spin pulse at the
orbital period suggesting the influence of accretion by a stream.
In addition, the presence of $3\omega$-2$\Omega_0$ and the other side-band 
components as listed in Table 2 supports the stream-fed
accretion for TV~Col.

The power spectra of TV~Col, thus, suggest a rather complex mode of accretion
in which the white dwarf accretes via both the disk and the
accretion stream. An asymmetry in the geometry of the X-ray source is
also implied. Similar power spectra
with different positive and negative side-band frequencies corresponding to
the spin and the orbital period are also observed in an IP 
TX Columbae by Norton et al. (1997), suggesting that TV~Col may have a close
resemblance with TX~Col.  However, unlike TX~Col, we did not observe any
change in the accretion mode for TV~Col during the two epochs of observations.

\subsection{X-ray Spin Modulation}

The spin modulation of the source is observed to be
nearly sinusoidal and the amplitude of the modulation is found
to decrease with increase in energy (Figs. 7, 8 \& 9).
The energy-dependent amplitudes have been measured more precisely here 
(see \S 3.3) as compared with the previous measurements with the $EXOSAT$ 
(Norton \& Watson 1989). The average weighted values of modulation 
depth reported by Norton \& Watson (1989) from three $EXOSAT$ observations
were $\sim$35\%$\pm$22\%, 31\%$\pm$3\%, 19\%$\pm$3\%, and 18\%$\pm$6\%
in the energy bands of 0.05--2.0 keV, 2--4 keV, 4--6 keV, and 6--10
keV respectively. Norton \& Watson (1989) defined the modulation 
depth as the peak-to-peak sinusoidal amplitude divided by the maximum 
flux and the values reported here are consistent with their observations. 
They also indicated the possibility of spin modulation in soft X-ray 
band being 180$^{\circ}$ out of phase compared with the hard X-ray band, 
as has been observed here in the $ROSAT$ and the $RXTE$ data.  
However, the energy-resolved folded light curves obtained from $ASCA$
vary in phase with each other (see \S 3.3).
The absence of any phase shift
in the two spin-folded $ASCA$ light curves, in the energy bands above 
and below 2 keV,  
suggests that the phase mismatch between the $ROSAT$ and the $RXTE$ 
light curves is most likely due to the lack of precision in the ephemeris for
the spin period of the white dwarf in TV~Col. 
Therefore, a precise measurement of the 
ephemeris for the spin of the white dwarf is required.

Energy dependence of the spin amplitude is one of the unique 
properties of IPs that is commonly observed in this subclass of 
magnetic CVs (e.g., in FO~Aqr by Beardmore et al. 1998). 
The change in 
the shape of the X-ray light curves with energy, from a sinusoid
in the soft X-rays to a near flat-top shape in the hard X-rays,
suggests that the harder X-ray emission comes from a relatively larger
region.
Since the most probable site for hard X-ray emission is the shocked
plasma in the post-shock region in the accretion column, therefore it 
is quite possible that in this case the shocked plasma is at a
considerable height above the white dwarf surface.
The hardness ratios HR1 and HR2 from the 1996 $RXTE$
data are spin-modulated but out of phase with the intensity 
variations.
The anti-correlation between the intensity and the hardness ratio 
curves HR1 and HR2 observed in TV~Col suggests that 
photoelectric absorption of X-rays above 2 keV could be responsible
for the modulation.
However, the lack of variation in the hardness ratio HR3 and a strong 
intensity modulation observed in the very soft band with the $ROSAT$ 
suggest that the softest X-rays are coming from an unobscured
region or that the absorbers responsible for variations above 2 keV
do not cover the source completely.
Below we test this hypothesis and try to estimate the amount of
absorption and covering fractions required to produce the observed 
spin modulation in TV~Col.

Broad-band X-ray spectrum taken with $ROSAT$ (averaged for the
1993 observation) and $RXTE$ (averaged for the longest observation
of 1996 August 13 with a similar modulation depth as the averaged data 
from all the 1996 observations) can be explained 
by a joint fit to a simple spectral model consisting of a partially but
heavily absorbed thermal bremsstrahlung component, interstellar absorption
in the line of sight and a Gaussian line at 6.6 keV.  The partial
covering model was successfully employed by Norton \& Watson (1989) 
to explain the $EXSOAT$ data of IPs, including TV~Col.  
If the bremsstrahlung continuum intensity is not constrained to be 
the same in the two observations separated by 3 yr, then the best-fit 
values for spectral parameters are as follows: 
covering fraction, C$_f$=0.94$^{+0.01}_{-0.04}$; 
column density, $N_H=7.5^{+0.6}_{-0.4} \times 10^{22}$ cm$^{-2}$;
thermal bremsstrahlung temperature, kT=23.5$\pm$1.2 keV; and
an interstellar absorption of 2.1$^{+0.2}_{-0.1} \times 10^{20}$ cm$^{-2}$ 
being consistent with other estimates in this direction. 
The minimum reduced $\chi^2$ ($\chi^2_{\nu}$) for this best-fit is 1.15 for 87 
degrees of freedom (dof) and errors quoted are with 90\% confidence.
Constraining the bremsstrahlung continuum to be the same in the two observations
gives the best-fit with $\chi^2_{\nu}$=1.23 for 88 dof and the 
best-fit parameters in this case are C$_f$=0.84$^{+0.01}_{-0.01}$,
$N_H=9.6^{+0.3}_{-0.4} \times 10^{22}$ cm$^{-2}$, and 
kT=21.7$^{+1.2}_{-1.0}$ keV, with the interstellar absorption being the
same as before.   In either case, keeping the values of the temperature 
and the Gaussian line emission parameters fixed at their best-fit values, 
we were able to reproduce the observed intensity and hardness ratio 
modulations by varying the $N_H$ and C$_f$. 
The value of $N_H$ had to be changed by 12\% -- 19\% from its best-fit 
value to reproduce the observed modulation in HR1 and HR2 and the count 
rates in the 2--5 keV, 5--10 keV and 10--20 keV energy bands.  
As the $N_H$ increases, HR1 and HR2 increase and the count rates above 
2 keV fall without affecting HR3 or the counts below 2 keV. 
A small simultaneous decrease ($\sim$ 0.06) in the C$_f$ is required  
to explain the simultaneous increase in the soft X-rays below 2 keV. 
A similar exercise, was also carried out for the  
$ASCA$ data using the same spectral model, as before.
It was observed
that a change in $N_H$ ($5.0^{+0.8}_{-0.7} \times 10^{22}$ cm$^{-2}$ 
for the averaged data) alone is not sufficient to reproduce the 
intensity modulation at energies below 2 keV, as was the case with the
$ROSAT$ data. In order to reproduce the intensity
modulations in the two $ASCA$ bands and the variations in HR4 we had to change
the value of $N_H$ by a factor of 2 accompanied by a simultaneous
increase (15\%) in $C_f$ ($0.69\pm0.02$ for the averaged data). 
In this case the best-fit values for the 
$N_H$ and $C_f$ are much lower than the corresponding values obtained from the $ROSAT$
and the $RXTE$ joint fit. Even though the   
change in the values of $N_H$ and $C_f$ required to reproduce the amplitude
modulation in $ASCA$ data is comparatively larger, it should be noted
however, that the maximum values of $N_H$ and $C_f$ are comparable with the
values required for the $ROSAT$ and the $RXTE$ data. These 
results suggest that a source 
partially covered with dense neutral material can successfully
explain the 
observed modulation in intensity and the hardness ratio curve in the 
$ASCA$ data also.

Therefore, it is quite possible to explain the observed
spin-modulated intensity and the hardness ratio curves by invoking a dense
partial absorber with varying hydrogen column density and a slightly
varying covering fraction in the accretion channel.
The real situation can be far more complex, however, with several ionized
and cold absorbers in front of shocked plasma emission component
as has been seen in higher spectral resolution data obtained with
$ASCA$ (Ezuka \& Ishida 1999).
Another possibility that the observed characteristics of the spin
modulation are due to a change in the shape of X-ray beam with energy
cannot be ruled out.

\subsection{X-ray Orbital Modulation}

The orbital modulation reveals light curves that are much more
structured and non-sinusoidal as compared with the spin-modulated light
curves. The hard X-ray (greater than 2 keV) intensity curves show a broad
energy-dependent dip between $\phi$=0.75--1.00.
Such energy dependent features modulated over orbital period are common
in IPs (Hellier et al. 1993).
Again in this case the energy dependent nature of this dip and its 
anti-correlation with hardness ratio suggests that the
photoelectric absorption is very likely the primary cause for its 
origin. 
The hardness ratios during this broad intensity dip are observed to 
increase significantly, thus indicating a strong absorption of
X-rays. Material with a high column density fixed in the
orbital frame is, therefore, required to cause these absorption 
effects. Repeating the exercise described in \S 4.2
we found that increasing the value of $N_H$
by 60\%--65\% from its best-fit value for the averaged data 
can reproduce the maximum values in HR1 and HR2 and the dip in the
count rates above 2 keV in $RXTE$ data.  
For the $ROSAT$ soft X-ray band the dip is the broadest (see Fig. 10) 
with a significant substructure but the corresponding hardness ratio HR3
(Fig. 11) does not clearly show a significant variation, 
as seen in HR1 and HR2.  There is, however, a data gap during 
this phase coverage followed by 
a small increase in HR3 during $\phi$=0.9--1.0 
accompanied by a decrease in intensity in the soft (less than 2 keV) 
X-ray band thus
giving some indication about the contribution from absorption effect 
to produce the dip.  The change in N$_H$ invoked above for explaining the 
$RXTE$ data in the dip is, however, not sufficient to
account for a corresponding change in the soft X-ray intensity during 
the dip, unless it is accompanied by a small increase ($\sim$0.02) in C$_f$.  
Similar results are obtained from the $ASCA$ data where 
increasing the value
of $N_H$ by $\sim$70\% from its average value can reproduce the observed
intensity minimum in the 2--10 keV band but is not sufficient to account for
the intensity minimum in the 0.7--2 keV band. In order to reproduce the 
dip minimum in both the energy bands and the maximum in the hardness ratio, we
have to simultaneously increase the value of $C_f$ by 20\%.
These results confirm that a partially covered absorber(s) can 
produce the dip modulated with the orbital period. It is also very likely that
the absorbing material is distributed non-homogeneously because of
the structured profile seen during the broad dip in the 
light curves and the hardness ratio curves (see Figs. 10, 11 \& 12).
According to a globally accepted view about orbital modulation in IPs,
this absorbing material is most likely situated near the site where the
accretion stream from the secondary impacts on to the accretion disk or
the magnetosphere. If this is the case with TV~Col then the inclination
of the system should be high enough for the material, splashed
out due to the impact of accretion stream, to cross the line of sight
and produce the absorption dip modulated over orbital period.
This is consistent with the estimated value of $70^{\circ}\pm3^{\circ}$
for the inclination angle of TV~Col obtained using the observed length and
depth of the eclipse from optical data (Hellier et al. 1991).
The broad dip in the hard X-ray light curves splits into two small 
dips (Fig. 10 \& 11)
suggesting that there may be two impact sites. This again implies that
the accretion stream is highly inhomogeneous.

Similar absorption dips at binary phase 0.7--0.8 were weakly hinted
at in the $EXOSAT$ observations reported by Hellier et al. (1993).
The absorption dips observed here with the $RXTE$ (and the $ASCA$) are clear, 
unambiguous and can be
seen in all the orbital cycles covered by the observations.
Such absorption
dips have also been observed in several other IPs and are similar to
those seen in the low-mass X-ray binaries (LMXBs) (Hellier et al. 1993).
The likely cause for these dips in LMXBs is considered to be absorbing
material above the plane of the accretion disk. In IPs where the stream
accretion is also present the impact of the stream with the magnetosphere
could produce shocks and throw some material out of the plane of the
accretion disk.

The light curve folded on the orbital period in the highest energy band
of 10--20 keV shows a negligible dip in intensity as compared with the 
folded light curves in the other energy bands. 
This suggests that the hard (10--20 keV) X-ray emission is almost
unaffected by the orbital motions of the secondary star, consistent with
the interpretation for dips at lower energies invoking absorbers.

\section{Conclusions}

Analysis of the longest hard and soft X-ray observations of TV~Col
with the $RXTE$, $ROSAT$ and $ASCA$ satellites leads us to the following
conclusions:

\begin{enumerate}

\item Power density spectra of X-ray emission from TV~Col show 
strong peaks at frequencies $\omega$ and $\Omega_0$ suggesting 
these to be two fundamental frequencies of the system, i.e., the
spin frequency and the orbital frequency.
The presence of spin modulation in X-rays is thus confirmed and
a significant power at $\omega$ is detected at all energies between 
0.1 to 20 keV. The $\Omega_0$ component
is clearly present in the energy bands below 10 keV, but is absent
in the hardest energy band of 10--20 keV.

\item Significant power at two fundamental periods $\omega$ and $\Omega_0$
and several side-band components of these two periods in the power
spectra of TV~Col provides evidence that contributions from both the
disk-fed and the stream-fed accretion are present.  
The presence of certain side-bands of $\omega$ suggests asymmetrically 
placed poles. The effects of 4-day precession period can also 
be seen in the side-bands of $\Omega_0$.

\item Modulation at the spin period being a function of energy is
confirmed. It is $\sim$ 3--4 times larger in the softest X-rays 
when compared with the value in the hardest X-rays observed. 
The observed energy dependent spin modulation and its anti-correlation 
with the hardness ratios defined above 2 keV can be explained by
a partially covered absorber system, with a small change in the
covering fraction being sufficient to explain the in-phase
soft X-ray modulation without affecting the X-ray intensity above
2 keV.

\item The effect of the orbital period is seen clearly in the hard 
and soft X-ray light curves of TV~Col.
Modulation due to the orbital period is also seen to be more
prominent in the soft X-rays than in the hard X-rays.  
X-ray light curves show a broad dip at orbital phase between 
$\phi$=0.7--1.0. The hardness
ratio variations during the dip can be understood by invoking 
an additional absorber (or absorbers) co-rotating in the orbital frame.
A significant substructure during the dip suggests that these additional 
absorbers may be inhomogeneous. The absorbers could be located near the region
where the accretion stream interacts with the accretion disk.

\end{enumerate}

\acknowledgments

This research has made use of data
obtained from the High Energy Astrophysics Science Archive Research
Center (HEASARC), provided by NASA's Goddard Space Flight Center.
VRR is pleased to acknowledge partial support from the Kanwal Rekhi
Scholarship of the TIFR Endowment Fund.
We thank an anonymous referee for useful comments and suggestions.
Starlink is funded by PPARC, and based at the Rutherford Appleton
Laboratory which is part of Council for the Central Laboratory of
the Research Councils, UK. The research of EMS was supported by 
contract number NAS8-39073 to SAO.  The research of PEB was supported 
by contract number NAG5-10247 to STScI.

%\clearpage

\begin{figure}
\plotone{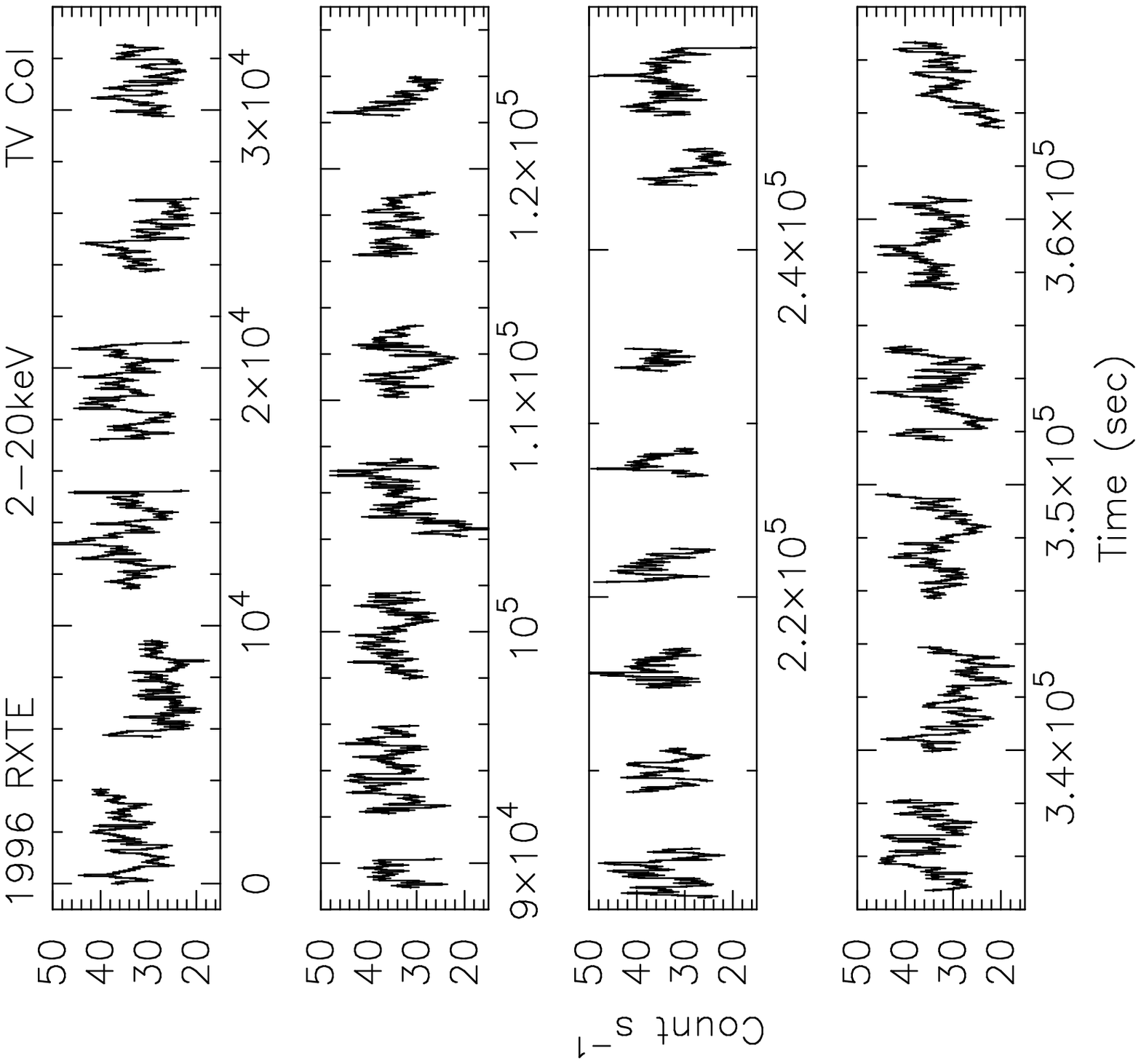}
\caption{Light curve of TV~Col in the 2--20 keV energy band obtained with $RXTE$
       during 1996 August 9--13. The bin time is 64 s for  the
       four panels. Time zero corresponds to TJD 10,304.5985. \label{fig1}}
\end{figure}

%\clearpage

\begin{figure}
\plotone{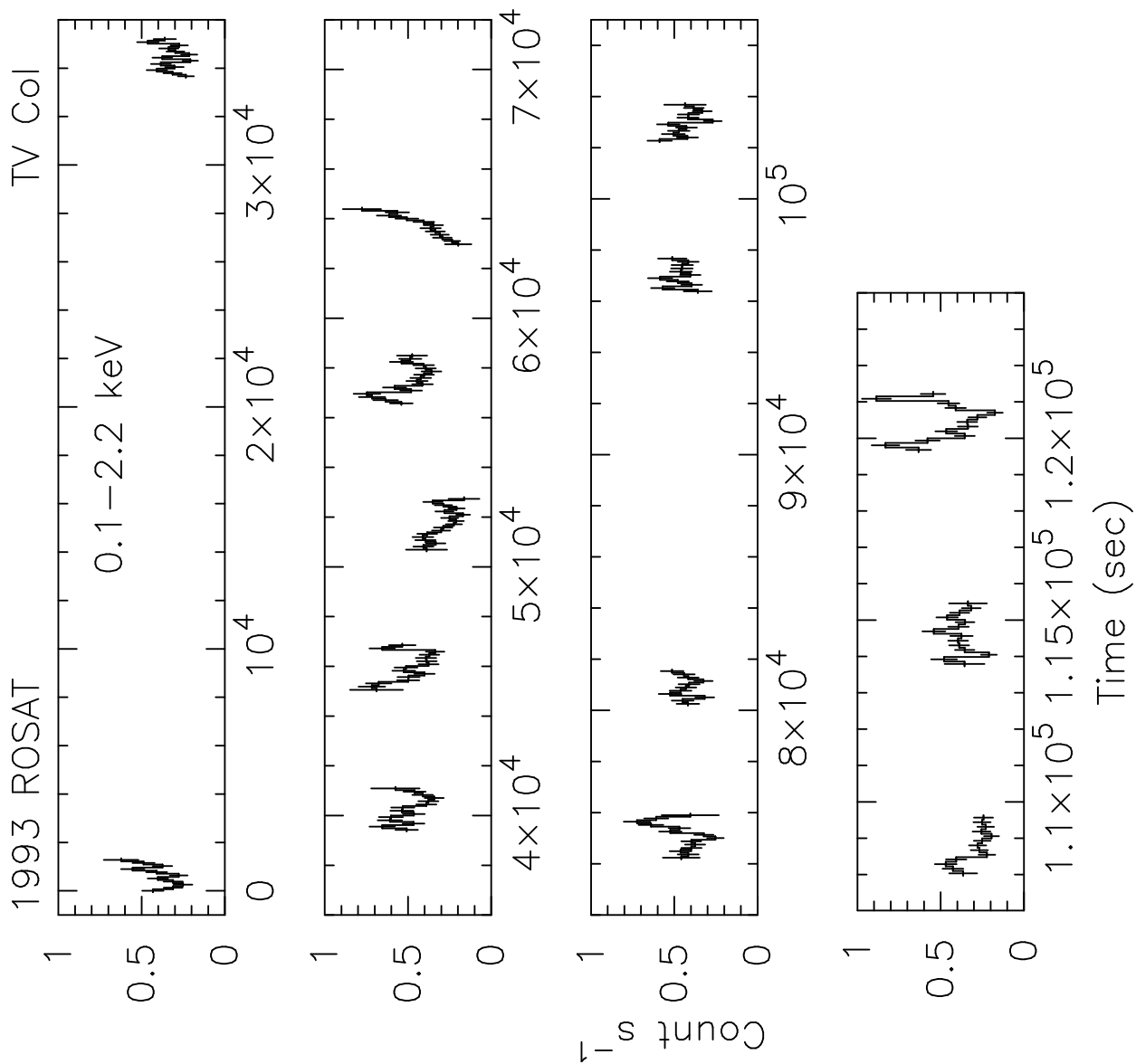}
\caption{Light curve of TV~Col in the 0.1--2.2 keV energy band obtained 
         with $ROSAT$
         during 1993 February 9--11. The bin time is 128 s for the
         four panels. Time zero corresponds to TJD 9,027.2597. \label{fig2}}
\end{figure}

%\clearpage

\begin{figure}
\plotone{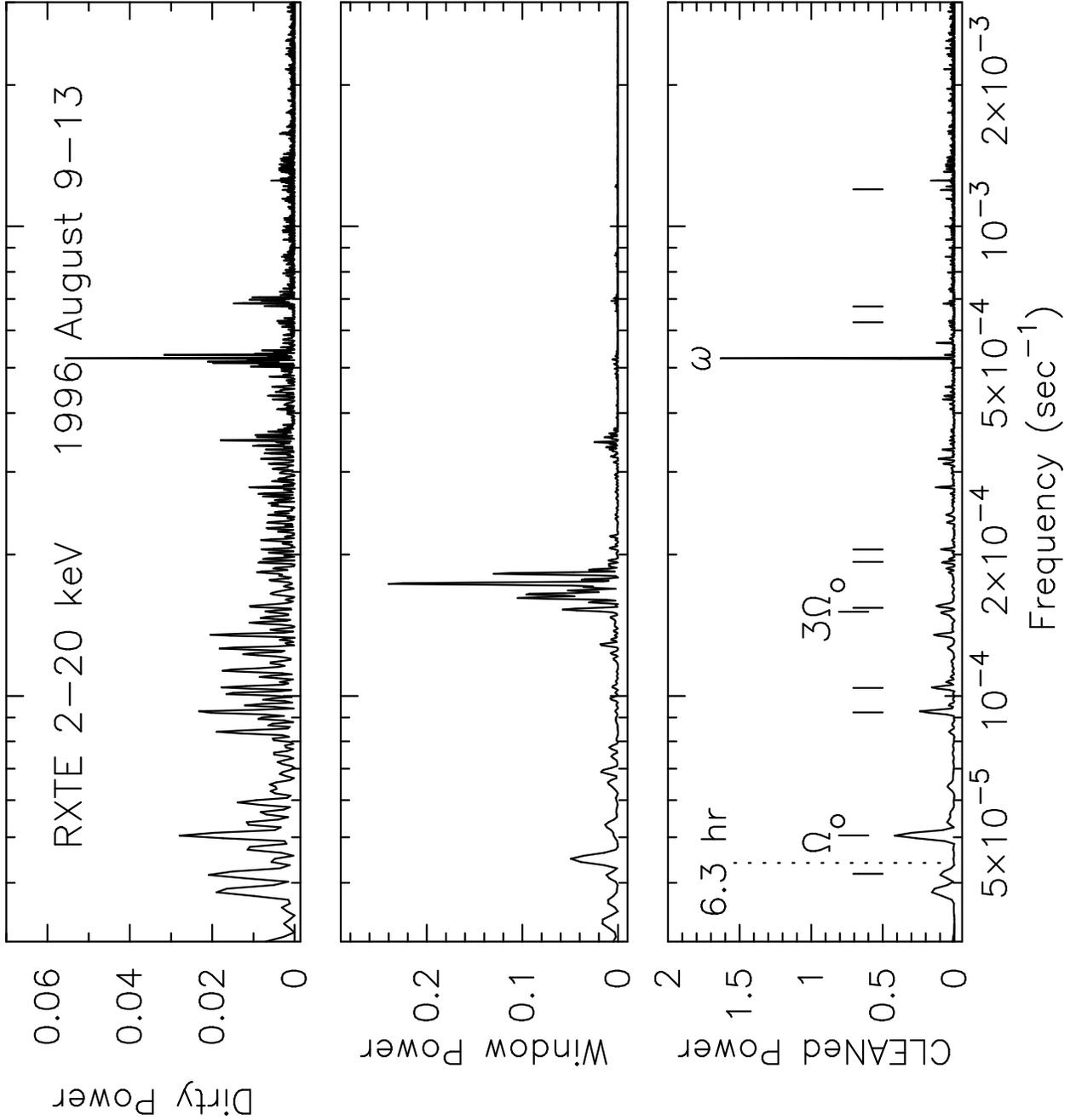}
\caption{Power spectra of TV~Col in the 2--20 keV energy
      band from the 1996 $RXTE$ observations. $Top$: a dirty
      power spectrum; $middle$, its corresponding window power. 
      $Bottom$: a CLEANed power spectrum for TV~Col.
      The tick marks show locations of various components that are detected and
      identified with components relevant
      to TV~Col at frequencies corresponding to the spin
      period ($\omega$), orbital period ($\Omega_0$), their side-bands, and
      also the side-band frequencies with
      precession period ($\Omega_{pr}$)(see Table 2). The location of 6.3 hr
      superhump period (Retter et al. 2003) is marked with a vertical dotted 
      line. \label{fig3}}
\end{figure}

%\clearpage

\begin{figure}
\plotone{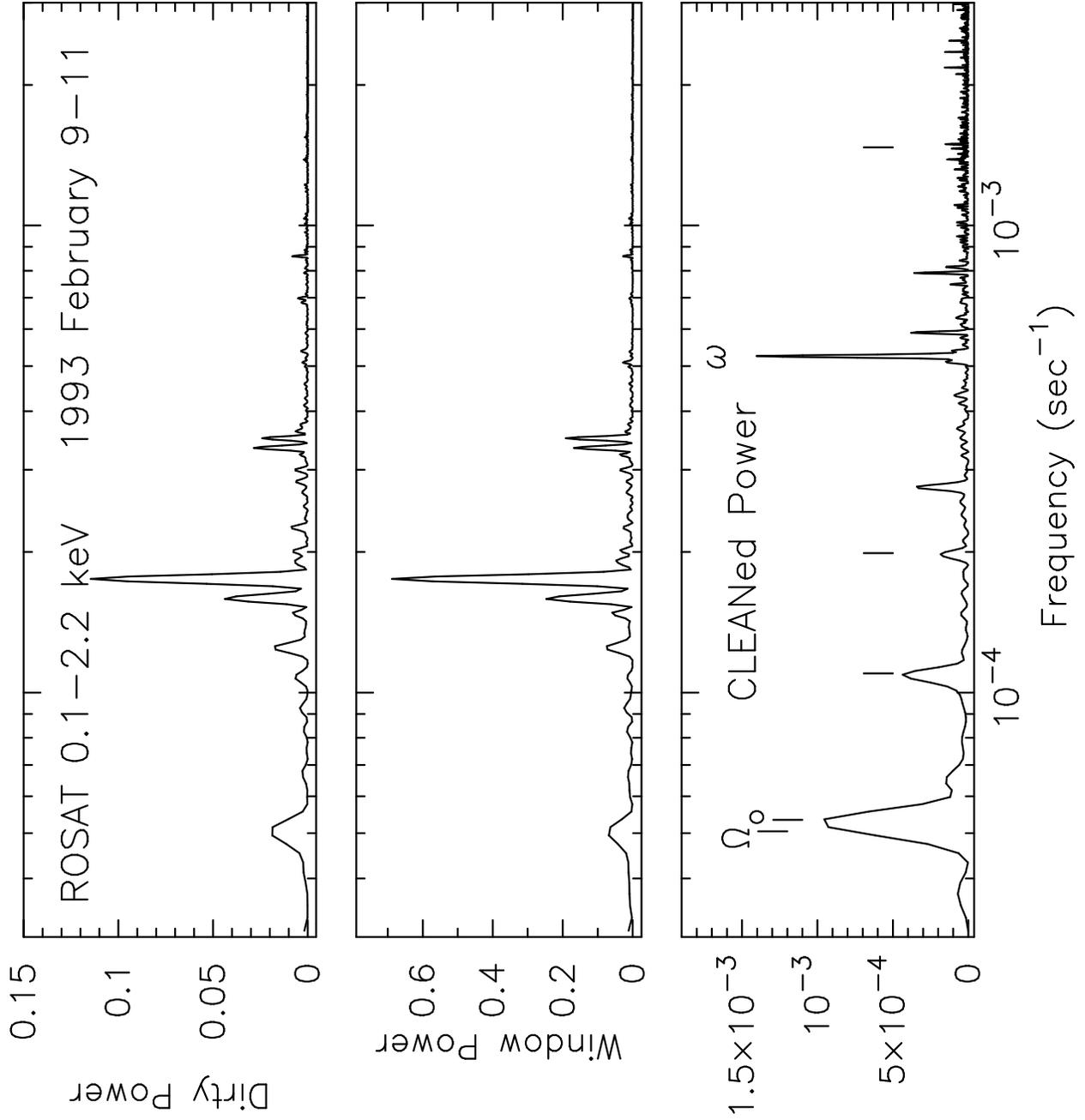}
\caption{Same as in Fig. 3, but for the 1993 $ROSAT$ observations
     in 0.1--2.2 keV energy band. Due to the poorer sampling
     and shorter length of observation, the peaks are not as well resolved
     here as in the $RXTE$ data.
     See Table 2 for the corresponding tick mark frequency
     components.\label{fig4}}
\end{figure}

%\clearpage

\begin{figure}
\plotone{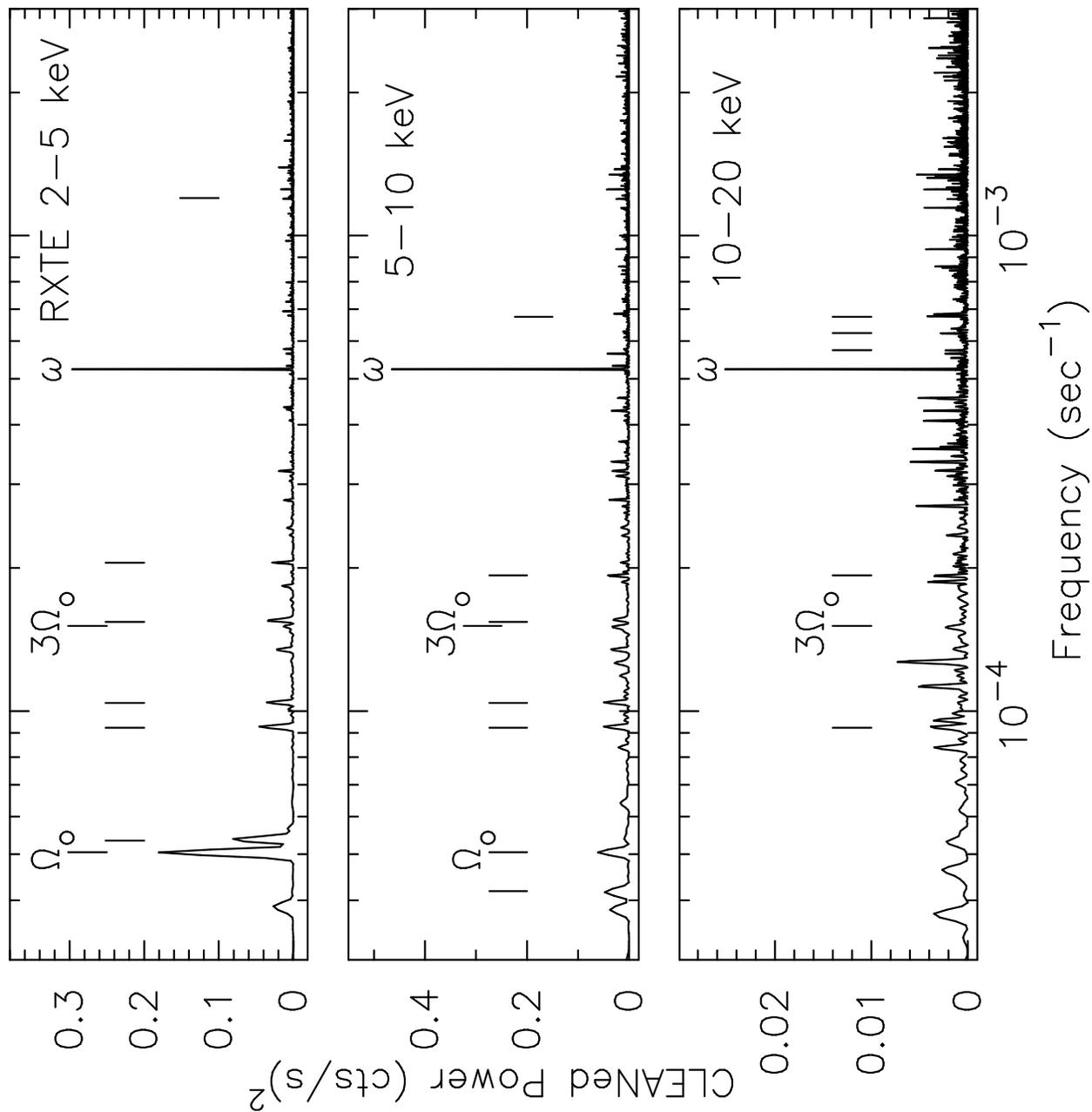}
\caption{Energy resolved CLEANed power spectra of TV~Col in 2--5 keV,
      5--10 keV, and 10--20 keV energy
      bands from the 1996 $RXTE$ observations. See Table 2 for the
      corresponding tick mark frequency components. \label{fig5}}
\end{figure}

%\clearpage

\begin{figure}
\plotone{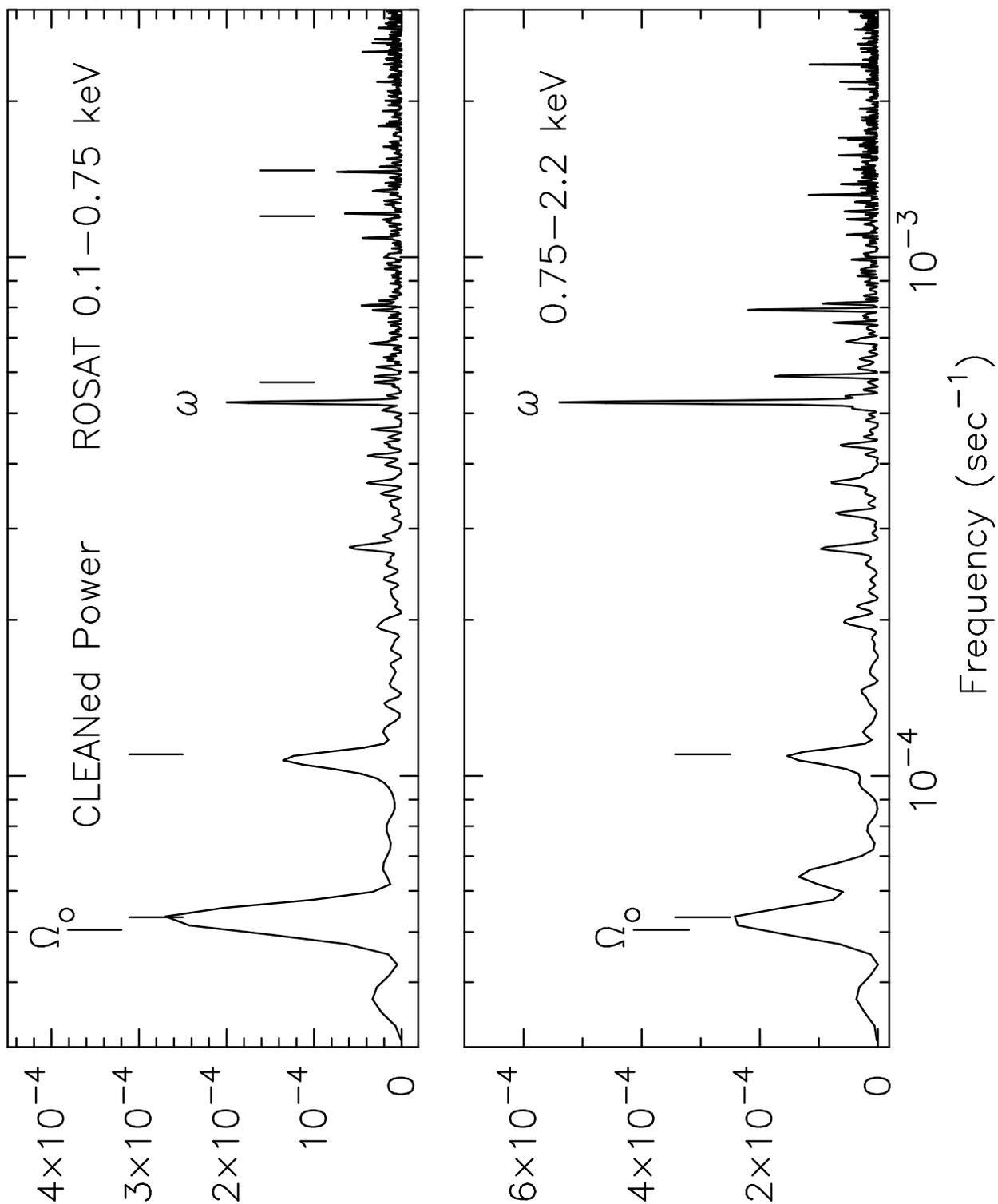}
\caption{Energy resolved CLEANed power spectra of TV~Col in soft X-ray
         energy bands of 0.1--0.75 keV and 0.75--2.2 keV
         from the 1993 $ROSAT$ observations. See Table 2 for the
         corresponding tick mark frequency components. \label{fig6}}
\end{figure}

%\clearpage

\begin{figure}
\plotone{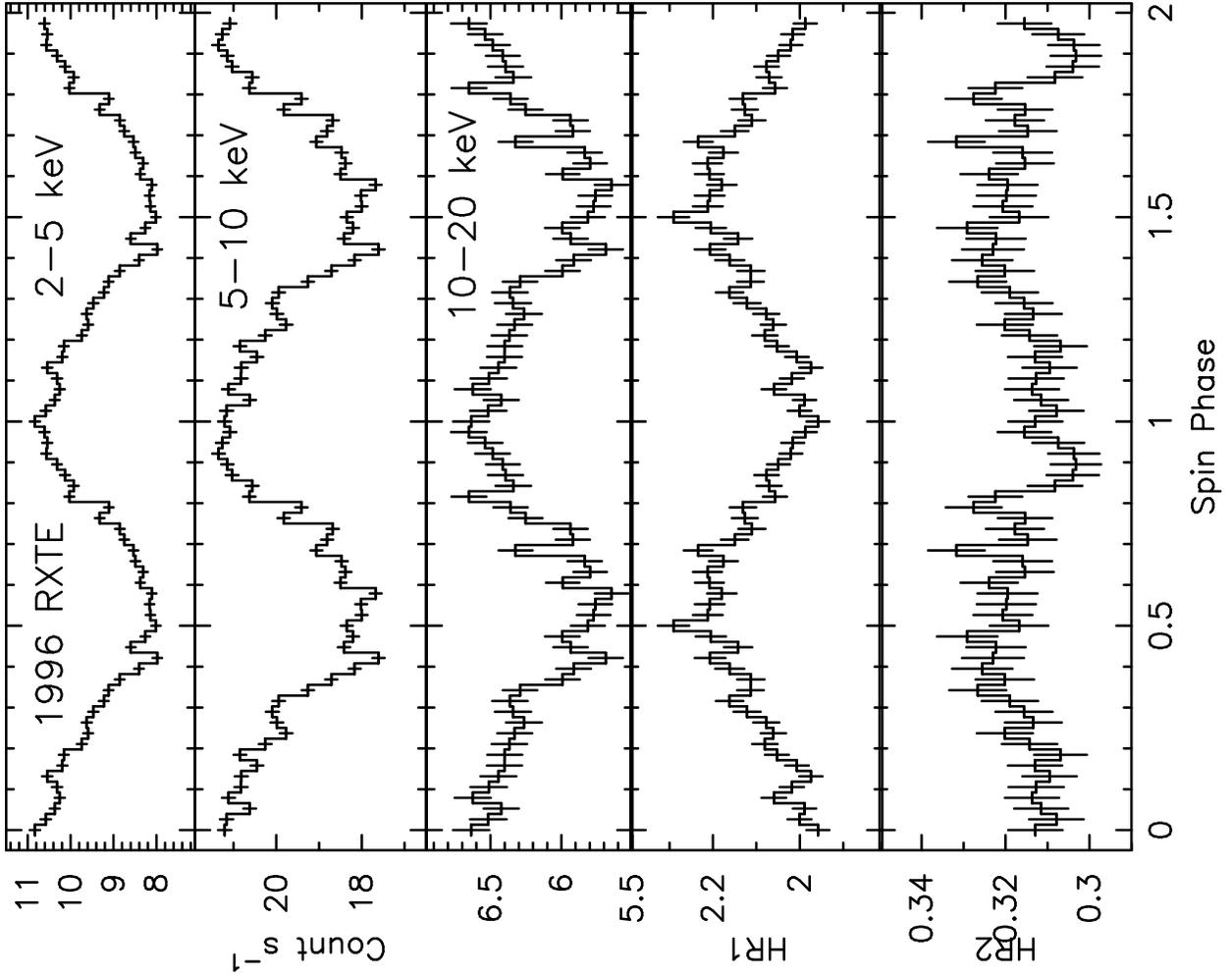}
\caption{Light curves folded on the spin
     period derived using the 1996 $RXTE$ data in 2--5 keV, 5--10 keV, and
     10--20 keV energy bands (top three panels). Hardness ratio
     curves HR1 and HR2 folded using the same spin period are shown in the
     bottom two panels (for definitions
     of HR1 and HR2 please see the
     text). The bin time is 50 s throughout. \label{fig7}}
\end{figure}

%\clearpage

\begin{figure}
\plotone{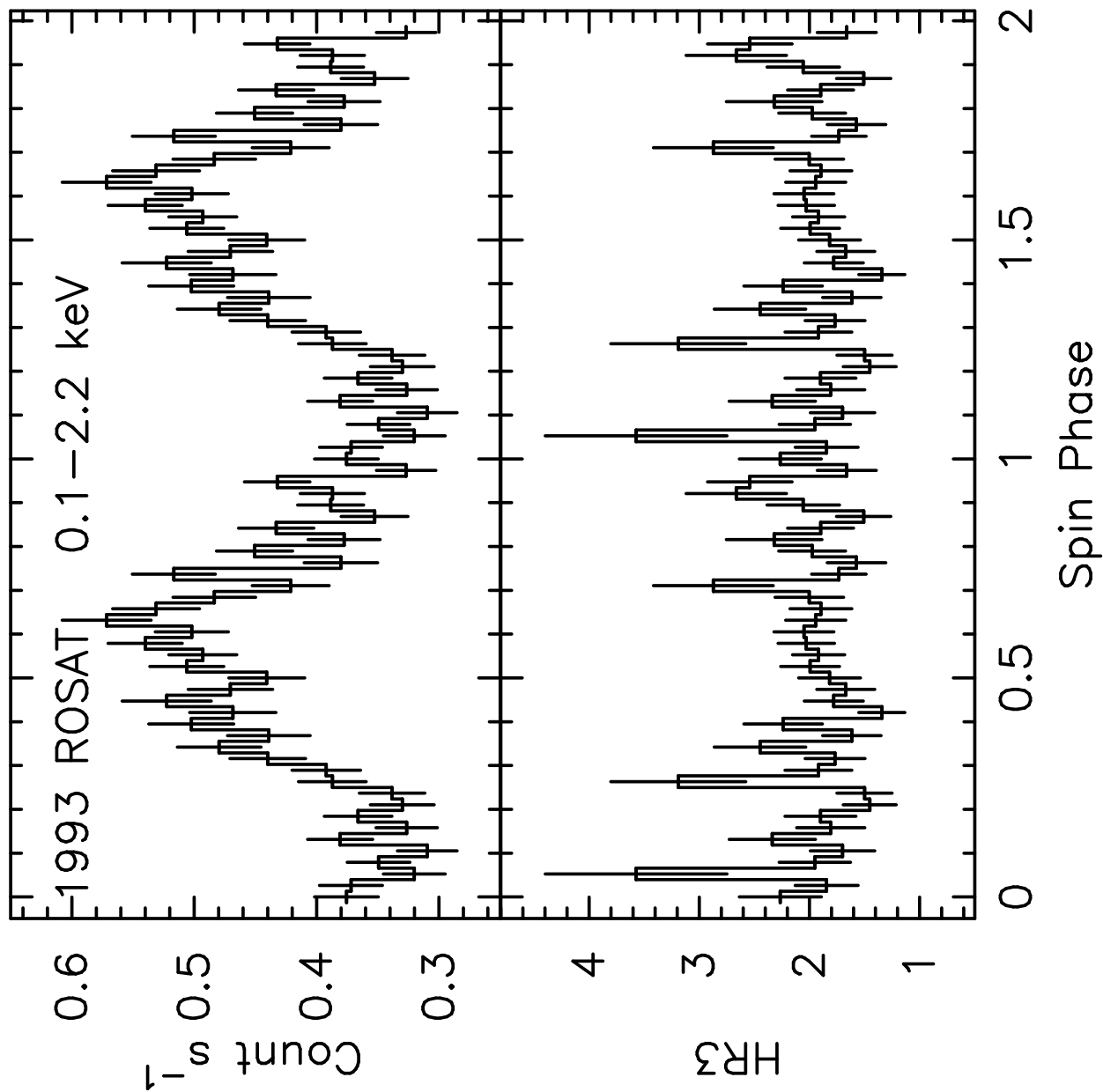}
\caption{Spin phase folded light curve in soft X-ray energy
     band of 0.1--2.2 keV
     from the 1993 $ROSAT$ observations ($top$). 
     $Bottom$: The hardness ratio curve HR3 which
     is defined in text. The bin time is 50 s throughout. \label{fig8}}
\end{figure}

%\clearpage

\begin{figure}
\plotone{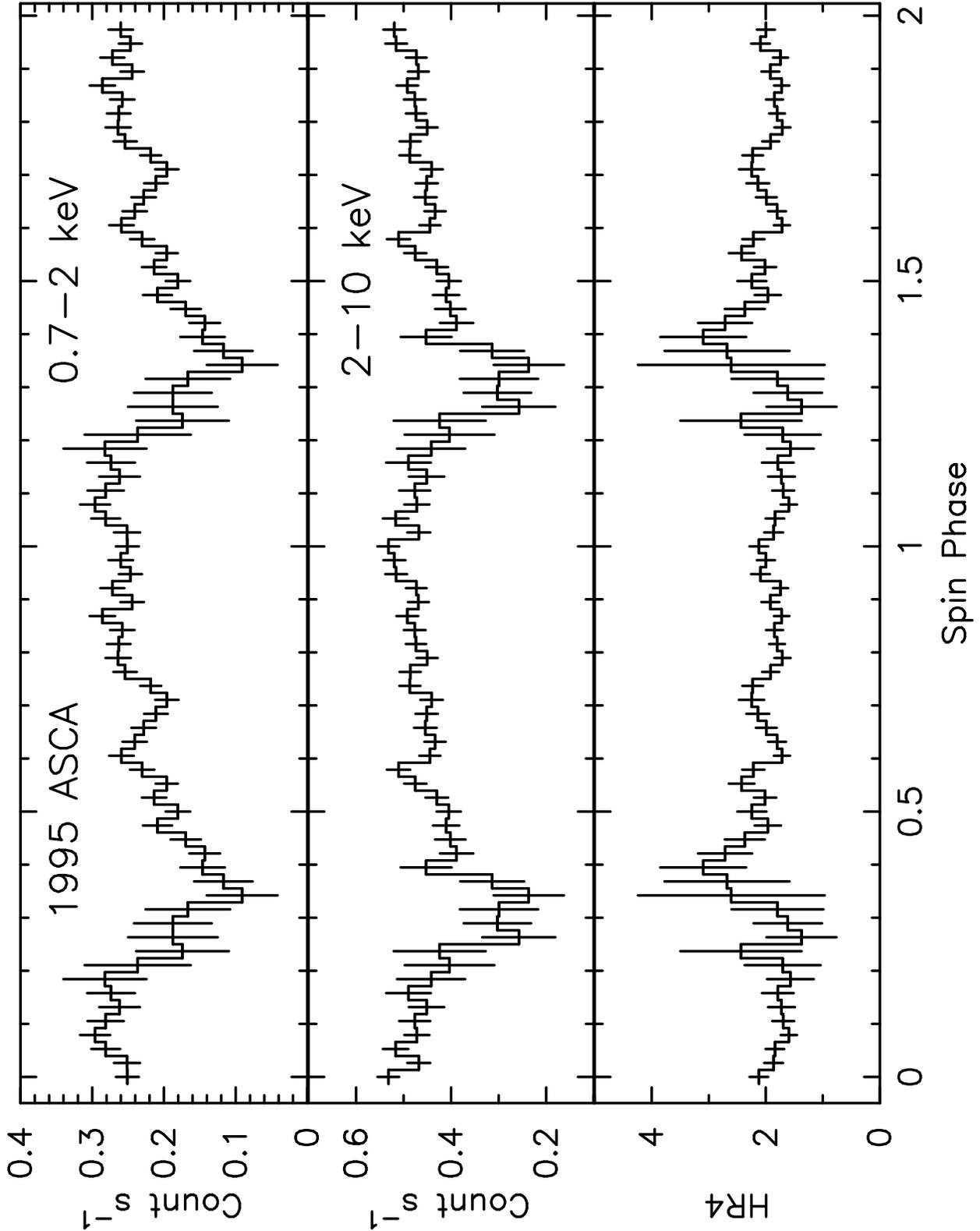}
\caption{Spin phase folded X-ray light curves in two energy bands from the 1995
ASCA SIS0 observations (top two panels). Hardness ratio curve HR4, as
defined in the text, is shown plotted in the bottom panel. The bin time is 50 s
throughout.  \label{fig9}}
\end{figure}

%\clearpage

\begin{figure}
\plotone{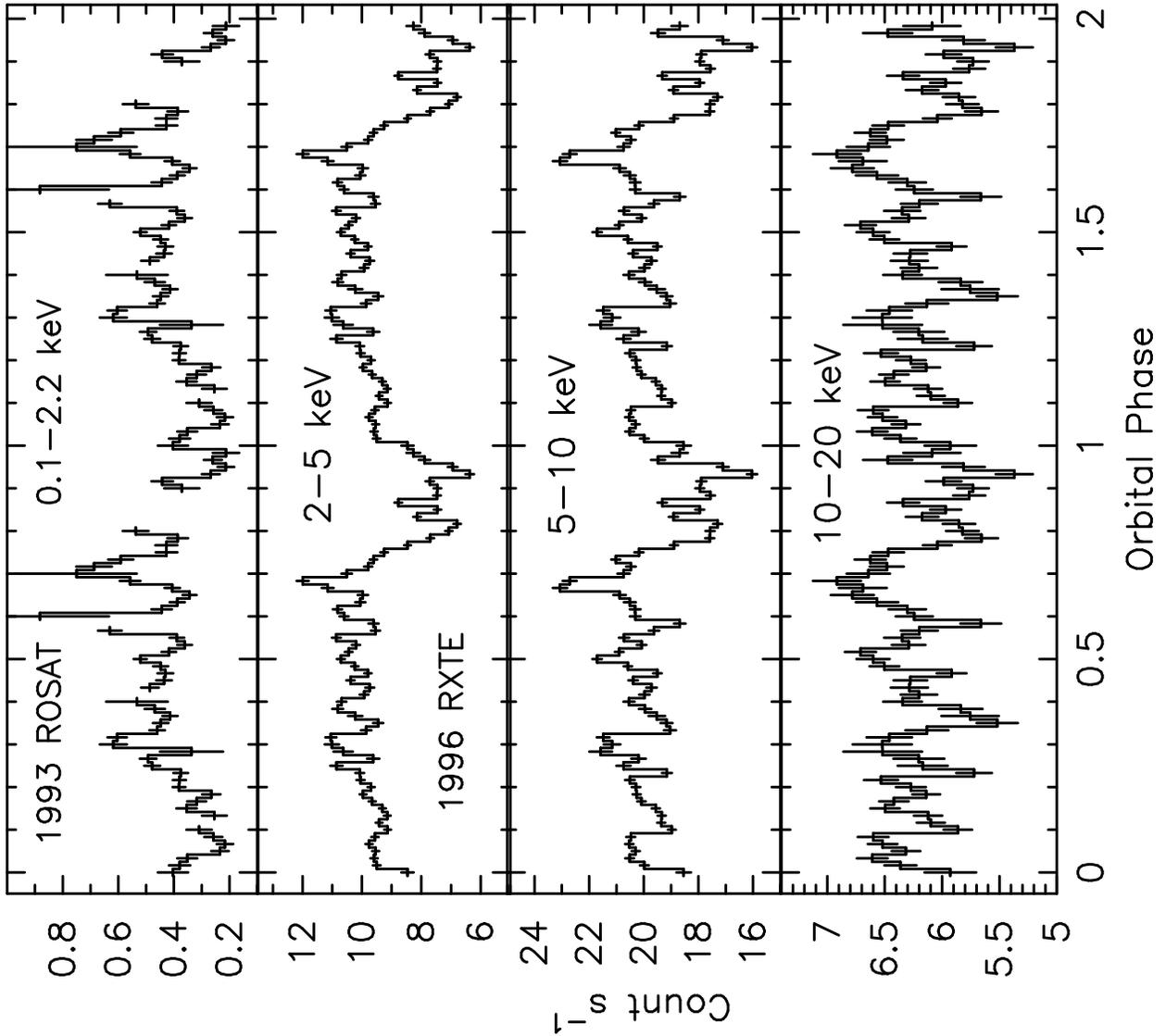}
\caption{Light curves folded on the orbital period using the
     ephemeris as given by Augusteijn et al. (1994).
     $Top$: the  $ROSAT$ soft X-ray light curve in
     0.1--2.2 keV energy band. The remaining three panels show $RXTE$ hard X-ray
     light curves in 2--5 keV, 5--10 keV and 10--20 keV energy bands,
     respectively. The bin time is 329~s throughout. \label{fig10}}
\end{figure}

\begin{figure}
\plotone{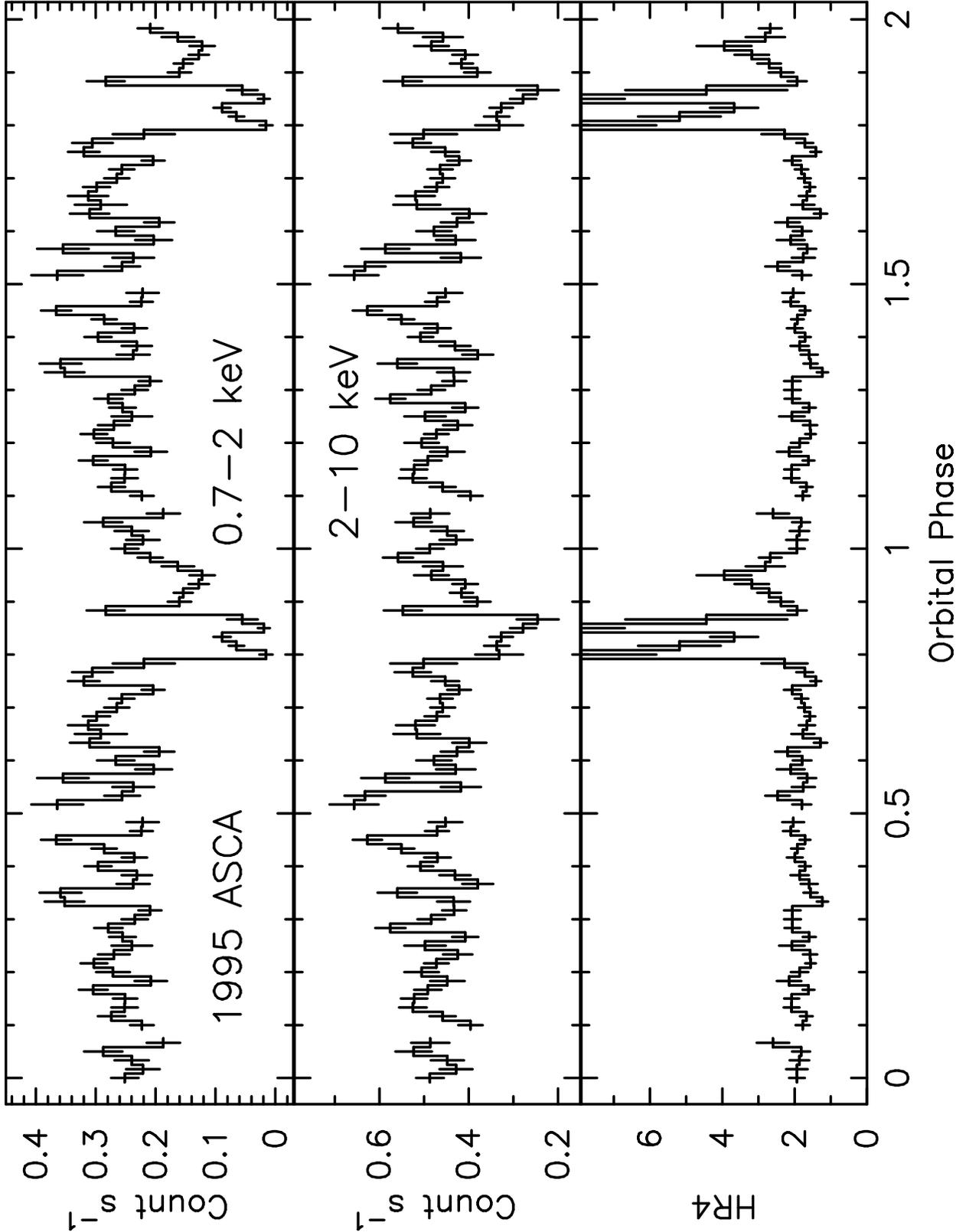}
\caption{Light curves folded on the orbital period using the ephemeris as
given by Augusteijn et al. (1994) from the 1995 ASCA SIS0 observations
(top two panels). Hardness ratio HR4, as defined in the text, is
shown plotted in the bottom panel. The bin time is 329~s throughout.
\label{fig11}}
\end{figure}

%\clearpage

\begin{figure}
\plotone{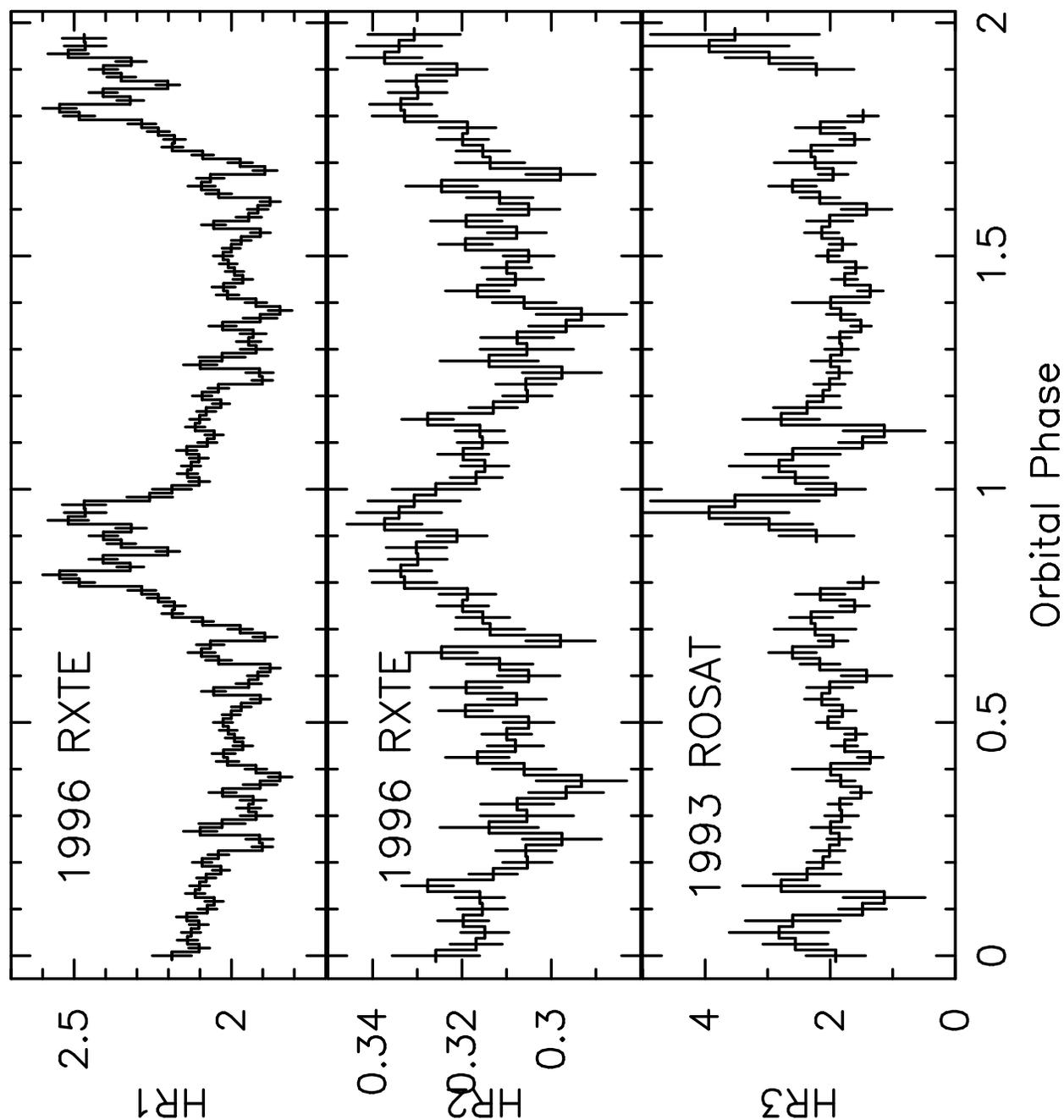}
\caption{Hardness ratio curves folded on the orbital period using the
     ephemeris as given by Augusteijn et al. (1994).
     For definitions of hardness ratios HR1, HR2, and HR3,
     please see the text. \label{fig12}}
\end{figure}

\clearpage

\begin{deluxetable}{cccccl}
\tabletypesize{\scriptsize}
\tablecaption{Summary of the $ROSAT$, $RXTE$, and $ASCA$ Observations of TV~Columbae. 
\label{tbl-1}}
\tablewidth{0pt}
\tablehead{
\colhead{Start Time} & \colhead{End Time} & \colhead{Satellite}   &
\colhead{Exposure} &
\colhead{Mean Count rate\tablenotemark{a,b}}  & \colhead{Comments}  \\
\colhead{(UT)} & \colhead{(UT)} &    &  
\colhead{ks}   & \colhead{(Count s$^{-1}$)}   &    \\
\colhead{(1)} & \colhead{(2)} & \colhead{(3)} & \colhead{(4)} &
\colhead{(5)} & \colhead{(6)}   
}
\startdata
1991 Feb 18, 0805... & Feb 20, 1530 & $ROSAT$ & 8.6 & 0.43 & Four exposures over \\
                  &          &         &     &      &  2 days, each for   \\
                  &          &         &     &      &  $\sim$2 ks  \\
1992 Sep 28, 0413... & Sep 30, 1551 & $ROSAT$ & 5.7 & 0.39 & Three exposures \\
                  &          &         &     &      & widely separated over 2\\
                  &          &         &     &      & days, each for $\sim$2 ks   \\
1993 Feb 9, 2009... & Feb 11, 0614 & $ROSAT$ & 22.6 & 0.43 & 14 exposures over   \\
                  &          &         &      &      & 2 days, each for  \\
                  &          &         &      &      & $\sim$2 ks \\
1996 Aug 9, 1422... & Aug 13, 2012 & $RXTE$  & 79 & 32.4 & Six to eight \\
     &     &     &   &    & exposures per day,   \\
     &     &     &   &    & each for 2 to 3.6 ks   \\
     &     &     &   &    & for a 4.2 day long   \\
     &     &     &   &    & observation  \\
2000 Dec 7, 0728... & Dec 09, 1229 & $RXTE$ & 22 & 33.0 & Two to four   \\
     &     &     &   &    & exposures per day,   \\
     &     &     &   &    & each for 2 to 3.6 ks   \\
     &     &     &   &    & for more than 2 days   \\
     &     &     &   &    & long observations  \\
1995 Feb 28, 0808... & Feb 01, 1040 & $ASCA$ & 24.7 & 0.58 & 26 exposures over   \\
     &     &     &   &    & $\sim$1 day, each for less    \\
     &     &     &   &    & than 2 ks   \\
\enddata

\tablenotetext{a}{The 0.1--2.2 keV energy band for the $ROSAT$, 2--20 keV energy band for the $RXTE$ and 0.7--10 keV energy band for the $ASCA$ satellite.}
\tablenotetext{b}{RXTE count rates have been normalized for 5 PCUs (see the text).}

\end{deluxetable}

%\clearpage

\begin{deluxetable}{lccccccccc}
\tabletypesize{\scriptsize}
\tablecaption{Absolute Power of Various Components. \label{tbl-2}}
\tablewidth{0pt}
\tablehead{
\colhead{} & \colhead{} & \multicolumn{8}{c}{ABSOLUTE POWER (Count s$^{-1}$)$^2$} \\
\\
\cline{3-10}
\\
\colhead{Frequency} & \colhead{Period}  & \multicolumn{3}{c}{$ROSAT$($\times$ 10$^{-4}$)} & \colhead{} & \multicolumn{4}{c}{$RXTE$}  \\
\cline{3-5} \cline{7-10}  \\
\colhead{Component} & \colhead{(sec)} & \colhead{0.1--2.2 keV} & \colhead{0.1--0.75 keV} & 
\colhead{0.75--2.2 keV} & \colhead{} & \colhead{2--20 keV} & 
\colhead{2--5 keV} & \colhead{5--10 keV} & \colhead{10--20 keV}   
}
\startdata
Noise \tablenotemark{a}  & \nodata & 0.4000 & 0.1500 & 0.2300 &  & 0.01  & 0.0012 & 0.0032 & 0.0008  \\
$\Omega_0$-3$\Omega_{pr}$  & 23909$\pm$394 & \nodata & \nodata & \nodata &  & 0.0989 & \nodata  & 0.0475 & \nodata     \\
$\Omega_0$          & 19819$\pm$167   & 9.2758 & 2.4250 & 2.3702 & & 0.4182  & 0.1805 & 0.0609 & \nodata    \\
$\Omega_0$+$\Omega_{pr}$   & 18727$\pm$235 & 9.5462 & 2.6936 & 2.4249 & & \nodata    & 0.0810 & \nodata  & \nodata   \\
2$\Omega_0$-3$\Omega_{pr}$ & 10831$\pm$79  & \nodata    & \nodata    & \nodata    & & 0.2408 & 0.0456 & 0.0494 & 0.0038 \\
2$\Omega_0$+$\Omega_{pr}$  & 9624$\pm$63   & \nodata    & \nodata    & \nodata    & & 0.1562 & 0.0357 & \nodata    & \nodata    \\
2$\Omega_0$+3$\Omega_{pr}$ & 9117$\pm$173  & 4.3506 & 1.3509 & 1.5346 & & \nodata    & \nodata    & \nodata    & \nodata    \\
3$\Omega_0$                & 6600$\pm$30   & \nodata    & \nodata    & \nodata    & & 0.0927 & 0.0128 & 0.0325 & 0.0023 \\
3$\Omega_0$+$\Omega_{pr}$  & 6476$\pm$30   & \nodata    & \nodata    & \nodata    & & 0.1250 & 0.0341 & 0.0309 & \nodata   \\
4$\Omega_0$-3$\Omega_{pr}$ & 5172$\pm$18   & \nodata    & \nodata    & \nodata    & & 0.0850 & \nodata    & 0.0118 & 0.0035 \\
4$\Omega_0$-$\Omega_{pr}$  & 5022$\pm$53   & 1.1842 & \nodata    & \nodata    & & \nodata    & \nodata    & \nodata    & \nodata    \\
4$\Omega_0$+$\Omega_{pr}$  & 4880$\pm$16   & \nodata    & \nodata    & \nodata    & & 0.0934 & 0.0285 & \nodata    & \nodata    \\
$\omega$              & 1909.67$\pm$2.5 & 14.038 & 1.9979 & 5.3934 &  & 1.6336 & 0.2966 & 0.4650 & 0.0252 \\
$\omega$+$\Omega_0$   & 1742$\pm$2   & \nodata    & 0.3095 & \nodata &   & \nodata     & \nodata    & \nodata    & 0.0023      \\
$\omega$+2$\Omega_0$  & 1601$\pm$2   & \nodata    & \nodata    & \nodata &   & 0.0322  & \nodata    & \nodata    & 0.0028 \\
$\omega$+3$\Omega_0$  & 1481$\pm$1.5 & \nodata    & \nodata    & \nodata &   & 0.0263  & \nodata    & 0.0108 & 0.0042 \\
2$\omega$+3$\Omega_0$ & 834$\pm$0.5  & \nodata    & 0.6458 & \nodata &   & 0.0943 & 0.0136 & \nodata    & \nodata    \\
3$\omega$-2$\Omega_0$ & 680$\pm$1    & 1.1100 & 0.7363 & \nodata &   & \nodata    & \nodata    & \nodata    & \nodata    \\
Unidentified & 8889$\pm$54 & \nodata & \nodata & \nodata &  & \nodata & \nodata & \nodata & 0.0051 \\
   & 7885$\pm$42 & \nodata & \nodata & \nodata &  & \nodata & \nodata & \nodata & 0.0073 \\
   & 7407$\pm$37 & \nodata & \nodata & \nodata &  & 0.1424 & 0.0218 & 0.0348 & \nodata \\
   & 3703$\pm$9  & \nodata & \nodata & \nodata &  & \nodata    & \nodata    & \nodata    & 0.0053  \\
   & 3650$\pm$27 & 3.3994 & 0.5926 & 0.9645 &  & \nodata & \nodata & \nodata & \nodata  \\
   & 3595$\pm$9  & \nodata & \nodata & \nodata &  & 0.1307 & 0.0126 & 0.0387 & \nodata  \\
   & 2993$\pm$6  & \nodata & \nodata & \nodata &  & 0.0818 & \nodata    & 0.0345 & 0.0059  \\
   & 2810$\pm$5  & \nodata & \nodata & \nodata &  & \nodata & \nodata & \nodata & 0.0057  \\
   & 2452$\pm$4  & \nodata & \nodata & \nodata &  & \nodata & \nodata & \nodata & 0.0046  \\
   & 2335$\pm$4  & \nodata & \nodata & \nodata &  & 0.0786 & \nodata & 0.0341 & 0.0045  \\
   & 2195$\pm$3  & \nodata & \nodata & \nodata &  & \nodata & \nodata & \nodata & 0.0051  \\
   & 1771$\pm$2  & \nodata & \nodata & \nodata &  & 0.1231 & \nodata & 0.0425 & 0.0023 \\
   & 1697$\pm$6  & 3.7944 & \nodata & 1.7409 &  & \nodata & \nodata & \nodata & \nodata \\
   & 1264$\pm$3  & 3.5933 & \nodata & 2.1954 &  & \nodata & \nodata & \nodata & \nodata \\
   & 800$\pm$0.4 & \nodata & \nodata & \nodata &  & 0.1642 & 0.0166 & 0.0433 & 0.0045 \\
   & 758$\pm$1   & \nodata & \nodata & 1.1718 &  & \nodata & \nodata & \nodata & \nodata  \\
   & 425$\pm$0.4 & \nodata & \nodata & 1.1595 &  & \nodata & \nodata & \nodata & \nodata  \\
\enddata

\tablecomments{Absolute power of various components detected above the noise 
level at frequencies corresponding to
the spin period ($\omega$ = 1909.67 s), orbital period ($\Omega_0$ = 5.5 hr)
and sideband frequencies from their combinations as predicted
in a model by Norton et al. (1996). Sideband
frequencies from the combination of orbital period and the 4-day
precession period ($\Omega_{pr}$), and strong but unidentified components
are also listed along with their power.}

\tablenotetext{a}{Absolute power for noise level with 90\% confidence.}

\end{deluxetable}


\begin{thebibliography}{}
\bibitem[Augusteijn(1994)]{Aug94} Augusteijn, T., Heemskerk, M. H. M., Zwarthoed, G. A. A., \& van Paradijs, J. 1994, A\&AS, 107, 219
\bibitem[Barrett(1988)]{Bar88} Barrett, P., O'Donoghue, D., \& Warner, B. 1988, MNRAS, 233, 759
\bibitem[Beardmore(1998)]{Bea98} Beardmore, A. P., Mukai, K., Norton, A. J., Osborne, J. P., \& Hellier, C. 1998, MNRAS, 297, 337
\bibitem[Bonnet(1985)]{Bon85} Bonnet-Bidaud, J. M., Motch, C., \& Mouchet, M. 1985, A\&A, 143, 313
\bibitem[Bretthorst(1988)]{Bre88} Bretthorst, G. L. 1988, Bayesian Spectrum Analysis and Parameter Estimation, (New York: Springer-Verlag)
\bibitem[Charles(1979)]{Cha79} Charles, P. A., Thorstensen, J., Bowyer, S., \& Middleditch, J. 1979, ApJ, 231, L131
\bibitem[Cook(1978)]{Coo78} Cooke, B. A., Ricketts, M. J., Maccacaro, T., Pye, J. P., et al. 1978, MNRAS, 182, 489
\bibitem[O'Donoghue(2000)]{Don00}O'Donoghue, D. 2000, New Astronomy Reviews, 44, 45
\bibitem[Ezuka(1999)]{Ezu99} Ezuka, H., \& Ishida, M. 1999, ApJS, 120, 277
\bibitem[Gregory(1996)]{Gre96} Gregory, P. C., \& Loredo, T. J. 1996, ApJ, 473, 1059
\bibitem[Hellier(1991)]{Hel91} Hellier, C., Mason, K. O., \& Mittaz, J. P. D. 1991, MNRAS, 248, 5
\bibitem[Hellier(1993a)]{Hel93a} Hellier, C., Garlick, M. A., \& Mason, K. O. 1993, MNRAS, 260, 299
\bibitem[Hellier(1993b)]{Hel93b} Hellier, C. 1993, MNRAS, 264, 132
\bibitem[Hutchings(1981)]{Hut81} Hutchings, J. B., Crampton, D., Cowley, A. P., Thorstensen, J. R., \& Charles, P. A.  1981, ApJ, 249, 680
\bibitem[Ishida(1995)]{Ish95} Ishida, M. \& Fujimoto, R. 1995, in Cataclysmic Variables, ed. Bianchini, A., Della Valle, M., \& Orio, M. (Holland: Kluwer), 93
\bibitem[Jahoda(1996)]{Jah96} Jahoda, K., Swank, J. H., Giles, A. B., Stark, M. J., Strohmayer, T., Zhang, W., \& Morgan, E. H. 1996, Proc. SPIE Vol. 2808, 59
\bibitem[Mateo(1985)]{Mat85} Mateo, M., Szkody, P., \& Hutchings, J. 1985, ApJ, 288, 292
\bibitem[McArthur(2001)]{Art01} McArthur, B. E., Benedict, G. F., Lee, J., van Altena, W. F., Slesnick, C. L., Rhee, J., Patterson, R. J., et al. 2001, ApJ, 560, 907
\bibitem[Motch(1981)]{Mot81} Motch, C. 1981, A\&A, 100, 277
\bibitem[Norton(1989)]{Nor89} Norton, A. J., \& Watson, M. G. 1989, MNRAS, 237, 853
\bibitem[Norton(1992a)]{Nor92a} Norton, A. J., Watson, M. G., King, A. R., Lehto, H. J., \& McHardy, I. M. 1992a, MNRAS, 254, 705
\bibitem[Norton(1992b)]{Nor92b} Norton, A. J., McHardy, I. M., Lehto, H. J., \& Watson, M. G. 1992b, MNRAS, 258, 697
\bibitem[Norton(1996)]{Nor96} Norton, A. J., Beardmore, A. P., \& Taylor, P. 1996, MNRAS, 280, 937
\bibitem[Norton(1997)]{Nor97} Norton, A. J., Hellier, C., Beardmore, A. P., Wheatley, P. J., Osborne, J. P., \& Taylor, P. 1997, MNRAS, 289, 362
\bibitem[Patterson(2001)]{Pat01} Patterson, J. 2001, PASP, 113, 736
\bibitem[Retter(2003)]{Ret03} Retter, A., Hellier, C., Augusteijn, T., Naylor, T., Bedding, T. R., Bembrick, C., McCormick, J., \& Velthuis, F. 2003, MNRAS, 340, 679
\bibitem[Roberts(1987)]{Rob87} Roberts, D. H., Lehar, J., \& Dreher, J. W. 1987, ApJ, 93, 968
\bibitem[Schrijver(1985)]{Sch85} Schrijver, J., Brinkman, A. C., van der Woerd, H., Watson, M. G., King, A. R., van Paradigs, J., \& van der Klis, M. 1985, Space Sci. Rev., 40, 121
\bibitem[Schrijver(1987)]{Sch87} Schrijver, J., Brinkman, A. C., \& van der Woerd, H. 1987, Ap\&SS, 130, 261
\bibitem[Singh(2003)]{Sin03} Singh K. P., Rana, V. R., Mukerjee, K., Barrett, P., \& Schlegel, E. M. 2003, in ASP Conference Ser. on MCVs, IAU Coll. 190 (Capetown), Eds: M. Cropper and S. Vrielmann
\bibitem[Tr\"{u}mper(1983)]{Tr83} Tr\"{u}mper, 1983, Adv. Space Res., 2, 241
\bibitem[Vrtilek(1996)]{Vrt96} Vrtilek, S. D., Silber, A., Primini, F., \& Raymond, J. C. 1996, ApJ, 465, 951
\bibitem[Wynn(1992)]{Wynn92} Wynn, G. A., \& King, A. R. 1992, MNRAS, 255, 83
\bibitem[Yamashita(1997)]{Yam97} Yamashita, A., et al. 1997, IEEE Trans. Nucl. Sci., 44, 847 

\end{thebibliography}
\end{document}